\documentclass[12pt]{article}

\usepackage{graphicx}
\usepackage{amssymb}
\usepackage{amsmath}
\usepackage{color}
\usepackage{natbib}

\begin{document}

\title{Meeting yield and conservation objectives by balancing harvesting of juveniles and adults}

\date{}

\maketitle

\noindent
{\small\emph{Niklas L.P. Lundstr\"om{\small{$\,^{1}$}}},
\emph{Nicolas Loeuille{\small{$\,^{2}$}}},
\emph{Xinzhu Meng{\small{$\,^{1, 3}$}}},
\emph{Mats Bodin{\small{$\,^{1, 4}$}}},
\emph{\AA{}ke Br\"annstr\"om{\small{$\,^{1, 6}$}}}}\\

\noindent
{\scriptsize{$^{1}$}}
{\scriptsize Department of Mathematics and Mathematical Statistics,}
{\scriptsize Ume\aa{} University, 90187 Ume\aa{}, Sweden},
{\scriptsize{$^{2}$}}
{\scriptsize Sorbonne Universit\'es, UPMC Univ Paris 6, UPEC, Univ Paris Diderot, Univ Paris-Est Cr\'eteil, CNRS, INRA, IRD,  Institute of Ecology and Environmental Sciences-Paris (IEES Paris),}
{\scriptsize place jussieu, 75005 Paris, France},
{\scriptsize{$^{3}$}}
{\scriptsize College of Mathematics and Systems Science​,}
{\scriptsize Shandong University of Science and Technology, Qingdao 266590, China},
{\scriptsize{$^{4}$}}
{\scriptsize Department of Ecology and Environmental Science,}
{\scriptsize Ume\aa{} University, 90187 Ume\aa{}, Sweden},
%
%
{\scriptsize{$^{6}$}}
{\scriptsize Evolution and Ecology Program,}
{\scriptsize International Institute for Applied Systems Analysis, 2361 Laxenburg, Austria}.\\\\





\noindent
{\bf{Abstract}}\\
Sustainable yields that are at least $80\%$ of the maximum sustainable yield are sometimes referred to as pretty good yield (PGY). The range of PGY harvesting strategies is generally broad and thus leaves room to account for additional objectives besides high yield. Here, we analyze stage-dependent harvesting strategies that realize PGY with conservation as a second objective. We show that (1) PGY harvesting strategies can give large conservation benefits and (2) equal harvesting rates of juveniles and adults is often a good strategy. These conclusions are based on trade-off curves between yield and four measures of conservation that form in two established population models, one age-structured and one stage-structured model, when considering different harvesting rates of juveniles and adults. These conclusions hold for a broad range of parameter settings, though our investigation of robustness also reveals that (3) predictions of the age-structured model are more sensitive to variations in parameter values than those of the stage-structured model. Finally, we find that (4) measures of stability that are often quite difficult to assess in the field (e.g.~basic reproduction ratio and resilience) are systematically negatively correlated with impacts on biomass and impact on size structure, so that these later quantities can provide integrative signals to detect possible collapses.\\


\section*{Introduction}

Almost one third of the world's fished marine stocks are currently overexploited (FAO 2016).
Some fish stocks have even collapsed, with examples including the
Californian sardine (\textit{Sardinops sagax}, Clupeidae) fishery in the 1950s \citep{R82},
the Atlanto-Scandian herring (\textit{Clupea harengus}, Clupeidae) fishery in the late 1960s \citep{KR92},
the Peruvian anchovy (\textit{Engraulis ringens}, Engraulidae) fishery in the 1970s \citep{C76},
and the Northern cod (\textit{Gadus morhua}, Gadidae) fishery off the east coast of Canada in the 1990s \citep{H96,OHLMBED04}.
The large proportion of overexploited marine fish stocks underscore the importance of implementing
sustainable harvesting practices and for further improving modern fisheries-management methods.

Maximum sustainable yield (MSY) has long been a central concept in population ecology
\citep{smithandpunt2001, hilborn2007, mesnil2012}. While maximization of yield from harvested populations is economically desirable, there is a rich scientific literature that criticizes the MSY concept and highlights its
shortcomings, including the difficulty of correctly estimating MSY, the inappropriateness of
long-term yield maximization as the single management objective, and the practical difficulty
of accurately implementing the required level of harvesting effort \citep{smithandpunt2001}.
MSY has further been criticized for its inability to prevent the collapse of important fisheries
\citep{BH57, L97, ML05, H10}. As an example, Alaska's Bering Sea Pollock fishery declined in 2009, and
despite being known as a sustainable fishery which implements scientific recommendations, the
management has been criticized for considering mainly MSY \citep{M09}.

MacCall and Hilborn have introduced the concept pretty good yield
(PGY; Hilborn 2010) for sustainable yields that are at least $80\%$ of the MSY.
In contrast to MSY harvesting-management objectives,
PGY can be realized by a range of
harvesting strategies and therefore leaves room to account
for other desirable objectives in addition to the maximization of yield.
The added value that PGY offers will
depend on the extent to which the implemented harvesting strategies can
successfully account for other desirable objectives beyond yield.

The aim of this paper is to investigate to which extent PGY harvesting strategies can
simultaneously account for high yield and large conservation benefits. To increase the
chances that our conclusions are valid over a broad range of circumstances, we base
our study on two established population models.
The first is an age-structured model (henceforth \textit{age model})
that is commonly used for modeling fish populations and evaluating fishing strategies
\citep{Francis92, Punt94, Punt_etal_95, PuntHilborn97, H10}.
The second is a stage-structured consumer-resource population model (henceforth \textit{stage model})
that has been introduced by de Roos et al. (2008).
Both models are capable of describing a range of aquatic and terrestrial animal populations.
The age model belongs to a class of models that have a long history in fisheries science and that incorporates age-dependent fecundity, age-dependent survival, and density-dependent recruitment.
The stage model is derived from a fully size-structured counterpart with food-dependent growth, fecundity and maturation,
and accounts for feedbacks from resource depletion. In particular, it accounts for ontogenetic asymmetry, i.e., differential abilities of juveniles and adults to utilize available resources \citep{BOOK13}.
With the age-model and stage-model being two fairly distinct representatives of contemporary population models,
results on which they agree are likely to be fairly robust and results on which they differ are likely to be ones where careful description of the population ecology is important and may therefore differ from species to species.
Using separate independently developed models to investigate robustness of findings should be a reasonable strategy 
(Levins 1966).

We extend both models by introducing selective harvesting of juveniles and adults,
giving wide ranges of possible harvesting strategies with different consequences for yield and
conservation. While it is straightforward to quantify the yield of a harvesting strategy,
it is less obvious how the conservations benefits should be measured. Here, we consider four
different measures of conservation benefits: two measures that capture the direct
impacts on the harvested population (the impact on population biomass and the impact on the
population size structure) and two measures that capture the indirect risks of collapse due to changes
in population dynamics (resilience and the basic reproduction ratio).
We determine trade-offs between yield and conservation benefits by finding the so-called
Pareto-efficient front; the set of strategies that cannot simultaneously be improved upon in
both yield and conservation benefit. These trade-off curves allow us to assess how large conservation benefits can be gained while preserving PGY. Finally, we determine the relationship between the direct
impact measures and the indirect risk measures, with the idea that the former are likely to be
more easily observable in the field while the latter better reflect the risks of collapse.
Taken together,
our results show that there are large potential gains of using specific PGY harvesting
strategies over traditional MSY strategies.
Moreover, among PGY strategies,  the ones that include equal harvesting of adults and juveniles often allow the best compromises between conservation and yield.





\section*{Methods}
\label{sec:methods}

In this section we first present the two population models,
one age-structured and one stage-structured.
%
%
%
We extend both models by introducing selective harvesting of juveniles and adults,
giving wide ranges of possible harvesting strategies with different consequences for the realized yield and for conservation.
We next present our methods of stability analysis involving the impact measures and risk measures that we will use to evaluate different harvesting strategies.
Finally, we recall the concept of maximum sustainable yield (MSY) and the economic concept of Pareto-efficiency which we will use to determine trade off curves between yield and conservation.


\subsection*{The age model}

We adopt an age-structured population model that in different guises has been widely used when
modeling fish populations and evaluating fishing strategies
\citep{Francis92, Punt94, Punt_etal_95, PuntHilborn97, H10}.
The model incorporates density-\\
dependent recruitment in the form of a Beverton-Holt
stock-recruitment relationship with a tunable degree of random recruitment variability.
Natural mortality is assumed to be independent of age and time,
and age-specific harvesting is assumed constant over time.
The central elements of the model,
which are mainly derived from Hilborn (2010),
are described below.

We denote by $N_{a,t}$ the number of individuals of age $a$ in year $t$ and assume that individuals mature at age $a_{\text{mature}}$ after which they reproduce at an age-dependent rate proportional to their body size.
Individuals younger than $a_{\text{mature}}$ are considered juveniles,
while individuals older than or with age equal to $a_{\text{mature}}$ are considered adults.
For simplicity, we assume that the population is made up entirely of female individuals,
but as we show in the Appendix,
our results are unchanged with a standard Fisherian sex-ratio of $50\%$ females.
The age-dependent fraction of mature females ($m_a$) and their corresponding egg production ($f_a$) are given by
\begin{align}\label{eq:ass_mature}
m_{a} = \left\{
\begin{array}{l}
0 \quad \textrm{if}\; a \leq a_{\text{mature}} \\
1 \quad \textrm{if}\; a > a_{\text{mature}}
\end{array}\right.
\qquad
f_{a} = c \, s_a,
\end{align}
where $c$ is a positive constant that scales the fecundity rate
and $s_a$ is the mass of an individual at age $a$.
We adopt a von Bertalanffy (1957) 
growth curve to describe individual length as a function of age,
and assume that individual mass $s_a$ is proportional to the cube of individual length, i.e.
\begin{align}\label{eq:growth_curve_B}
s_a &=  s_{\text{max}} \left(1 - e^{-K(a - a_0)}\right)^{3},
\end{align}
where 
$s_{\text{max}}$ is the asymptotic maximum body mass,
$K$ is a growth rate parameter and
$a_0$ is a hypothetical negative age at which the individual has zero length.
In Appendix Fig A 3 we illustrate von Bertalanffy growth curves for some
parameter values. 

The total egg production in a year $t = 0,1,2, \dots$ is found by summing over the offspring produced by mature females of different ages,
\begin{align}\label{eq:egg_production_1}
E_t = \sum_{a = 0}^{a_{\text{max}}} m_a f_a N_{a,t},
\end{align}
%
where $a_{\text{max}}$ is the maximum age of individuals and $N_{a,t}$ is the number of individuals, 
per unit of volume, of age $a$ in year $t$.

The number of individuals in each age class changes from year to year according to
\begin{align}\label{eq:main_age_model}
&N_{0,t} = R_t, \qquad \text{and}\notag \\ 
&N_{a,t} = N_{a-1,t-1} S (1 - \gamma_{a-1})\qquad \text{for} \; 1 \leq a,
\end{align}
%
where $R_t$ is the recruitment of newborn individuals in year $t$, $t = 1,2,3,\dots$, as described further below,
and $S(1 - \gamma_{a-1})$ is the probability that an individual survives from one year to the next.
This survival probability is decomposed in survival from natural mortality $S$ and from harvesting mortality
$(1 - \gamma_{a-1})$.
Note that, as probabilities, these variables always take values from 0 to 1.

We incorporate stage-selective harvesting by allowing separate constant
fractions harvested of juveniles ($F_{\text{J}}$) and adults ($F_{\text{A}}$)
and setting the vulnerability of individuals to
\begin{align}\label{eq:harvest_rates_age-model}
\gamma_{a} = \left\{
\begin{array}{l}
F_{\text{J}} \quad \textrm{if}\; a < a_{\text{mature}} \\
F_{\text{A}} \quad \textrm{if}\; a \geq a_{\text{mature}}.
\end{array}\right.
\end{align}

We assume Beverton-Holt recruitment \citep{BH57},
\begin{align}\label{eq:recruitment}
R_{t+1} = \frac{E_{t}}{\alpha + \beta E_{t}} \text{exp} \left( u_t- \frac{\sigma_u^2}{2}\right), 
\end{align}
where $u_t$ are independent and normally distributed random variables with mean 0 and standard deviation $\sigma_u$.
The factor $-\sigma^2_u/2$ ensures that the expected number of recruits remains the same with varying $\sigma^2_u$
because the lognormal random variable represented by the exponential has always mean 1. ​
The parameters $\alpha$ and $\beta$ are
not used directly as they lack a direct ecological interpretation.
Instead, they are determined from the expected total egg production at equilibrium in absence of harvesting ($\mathcal{E}_0$)
and from the steepness parameter ($h$) setting the sensitivity of the recruitment to the total egg production.
The steepness is defined as the ratio of recruitment when egg production equals $20\%$ of $\mathcal{E}_0$ to recruitment at $\mathcal{E}_0$ \citep{MD88, H10} and may take values between 0.2 and 1.
If $h$ is close to 1, recruitment is almost independent of the egg production,
and if $h$ is close to 0.2, recruitment is almost proportional to the egg production.
In Appendix Fig. A 1 we illustrate how recruitment depends on $\mathcal{E}_0$ and $h$ and
state their exact mathematical relationship with $\alpha$ and $\beta$.

For the age model, we use
\begin{align}\label{eq:param_age}
\mathcal{R}_0 = s_{\text{max}} = c &= 1, \quad a_{\text{max}} = 100, \quad K = 0.23,  \quad a_0 = -2,  \notag\\ 
a_{\text{mature}} &= 8, \quad h = 0.7, \quad S = 0.8, \quad \sigma_u = 0,
\end{align}
%
%
%
%
%
as our default parameters values, with substantial motivations given in the Appendix.
Here, $\mathcal{R}_0$ is measured in number of individuals per unit of volume,  
$s_{\text{max}}$ and $c^{-1}$ have an arbitrary mass unit, 
while $a_{\text{max}}$, $a_0$, $a_{\text{mature}}$ and $K^{-1}$ are measured in years.
However, after a rescaling of the equations,
$\mathcal{R}_0$, $s_{\text{max}}$ and $c$ (as well as $s_a$, $E_t$, $R_t$ and $N_{a,t}$) 
can be considered as non-dimensional and we can take $\mathcal{R}_0 = s_{\text{max}} = c = 1$ without loss of generality. 
See the Appendix for details. 
%
While we believe that the parametrization in \eqref{eq:param_age} is a reasonable choice,
we have considered substantial variations and present how our results from the age model
depend on parameter values in the result section.
A systematic investigation of the robustness of our results with respect to variations of the parameter values in
\eqref{eq:param_age} are given in the Appendix.


\subsection*{The stage model}

We adopt an archetypal consumer-resource model that has been introduced by de Roos et al. (2008)
as a reliable approximation of a fully size-structured population model.
The model is stage-structured and incorporates key aspects of individual life history such as food-dependent growth, maturation, and fecundity.
In contrast to the age-model, the population-level feedback that results from resource depletion induces competition between life-history stages whenever juveniles and adults have differential abilities to utilize available resources.
Competition under such ontogenetic asymmetry can strongly influence the ecological dynamics \citep{BOOK13} and may thus effect how harvesting affects population size structure.
The central elements of this model are described below, with the detailed model formulation given in
de Roos et al. (2008).

Individuals are composed into two stages, juveniles and adults, depending only on their size.
Both juveniles and adults forage on a shared resource $R = R(t)$.
The juvenile biomass is denoted by $J = J(t)$ while adult biomass is denoted by $A = A(t)$.
Juveniles are born with size $s_{\text{born}}$ and grow until
they reach the size $s_{\text{max}}$ at which point they cease to grow,
mature, and become adults.
Juveniles use all available energy for growth and maturation,
while adults do not grow and instead invest all their energy in reproduction.
The juvenile growth rate and adult reproduction rate depend on resource abundance.
In accordance with metabolic theory of ecology,
foraging ability and metabolic requirements increase with individual body size \citep{BGASW04}.
Juveniles and adults do not produce biomass when the energy intake is insufficient to cover
maintenance requirements.

The rate at which the biomass of juveniles, adults, and available resources changes are given by three differential equations:
\begin{eqnarray}\label{eq:stage-model}
\frac{dJ}{dt} &=& \left( w_{\text{J}}(R) - v(w_{\text{J}}(R)) - M - F_{\text{J}}\right) J  + 
w_{\text{A}}(R)A,\nonumber \\
\frac{dA}{dt} &=& v(w_{\text{J}}(R))J - \left( M + F_{\text{A}}\right) A,\\
\frac{dR}{dt} &=& r(R_{\max}-R)
-I_{\max}\frac{R}{H+R}\left(J+qA\right).\nonumber
\end{eqnarray}
Here, $w_{\text{J}}(R)$ and $w_{\text{A}}(R)$ are the net biomass production, 
per unit body mass, of juveniles and adults, respectively.
The natural mortality is denoted by $M$ while $F_{\text{J}}$ and $F_{\text{A}}$ are the respective
stage-dependent harvesting rates of juveniles and adults.
Continuing, $v(w_{\text{J}}(R))$ is the resource-dependent rate at which juveniles mature and become adults,
$r$ is the resource turnover rate, and $R_{\max}$ is the maximum resource density.

The net biomass production rates for juveniles and adults are assumed to equal the balance between ingestion and mass-specific metabolic rate $T$ according to
\begin{equation*}
w_{\text{J}}(R)= \max\left\{ 0, \sigma I_{\max} \frac{R}{H+R} - T \right\} \quad \text{and}\quad
w_{\text{A}}(R)= \max\left\{ 0, \sigma q I_{\max} \frac{R}{H+R} - T \right\}.
\end{equation*}
Here, $\sigma$ represents the efficiency of resource ingestion, and the maximum juvenile and adult ingestion rates per unit biomass equal $I_{\max}$ and  $qI_{\max}$, respectively, $H$ is the half-saturation constant of consumers, and the factor $q$ describes the difference in ingestion rates between juveniles and adults.
The juvenile maturation rate depends on the net biomass ingestion and thus also on the resource density.
It is derived from the fully size-structured counterpart by assuming that the population size structure is at equilibrium and by determining the rate at which juvenile individuals reach the maturation size $s_{\text{max}}$.
In the stage model, the juvenile maturation rate is given by
\begin{align*}
v(w_{\text{J}}(R)) = \frac{ w_{\text{J}}(R) - M - F_{\text{J}}} {1-\left(s_{\text{born}}/s_{\text{max}}\right)^{1-(M + F_{\text{J}})/ w_{\text{J}}(R)}}
\end{align*}
The function $v(w_{\text{J}}(R))$ results from a mathematical derivation
and lacks a clear biologically interpretable form.
When $w_{\text{J}}(R) = M + F_{\text{J}}$ the function is undefined, and at this value it is defined by $v(M + F_{\text{J}}) = -(M + F_{\text{J}})/\log(s_{\text{born}}/s_{\text{max}})$.
In Appendix Fig. A 7 we illustrate the maturation function $v(w_{\text{J}}(R))$
as well as the net biomass production functions $w_{\text{J}}(R)$ and $w_{\text{A}}(R)$.
Noting that the size at birth $s_{\text{born}}$ and the size at maturation $s_{\text{max}}$ appears only as the
fraction $s_{\text{born}} / s_{\text{max}}$ 
we can reduce the numbers of parameters by letting $z = s_{\text{born}}/s_{\text{max}}$.

For the stage model, we use
\begin{align}\label{eq:param_stage}
H = T = r = 1, \quad R_{\text{max}}=2, \quad \sigma = 0.5, \quad I_{\text{max}} = 10, \quad M = 0.1, \quad z =0.01, \quad q = 0.85,
\end{align}
%
%
%
%
%
%
%
as our default parameters values.
Here, $H$ and $R_{\text{max}}$ are measured in biomass per unit of volume
while $T, r, I_{\text{max}}$ and $M$ are expressed per unit of time.
However, after a rescaling of the equations all parameters, 
as well as the biomass densities $J$, $A$ and $R$, 
can be considered as non-dimensional and we can take $H = T = 1$ without loss of generality. 
See the Appendix for details. 
The parameter values in \eqref{eq:param_stage} are inspired by 
de Roos et al. (2008) 
and may be considered as archetypal.
We show in the Appendix that our results are largely robust with respect to variations of these default values.


\subsection*{Model dynamics}

Both the age model and the stage model are nonlinear dynamical systems and therefore complicated dynamics can not be ruled out a priori.
However, extensive numerical investigations of basin of attractions indicate that solution trajectories end up,
after sufficient time, at a globally stable equilibrium in both models.
This equilibrium is therefore the only attractor which is either an interior (positive) equilibrium
or an extinction equilibrium depending on the harvesting rates.
Indeed, our nonlocal approach to resilience (presented below) tests the dynamics in both models for a large number
of perturbations (inferred through initial conditions) and would detect any coexisting attractor with a high probability.
We refer the reader to Meng et al. (2013) and  Roos et al. (2008) for more on dynamic properties,
mathematical analysis as well as numerical investigations of the stage model.


\subsection*{Stability analysis: measures of conservation}

Stability of ecological systems is important for both conservation and harvesting purposes. In unstable systems, population dynamics may transiently go to low biomass values where the populations become vulnerable to demographic stochasticity or other factors. Hence, lack of stability promotes extinction. Stability is desirable also for harvest managers as it ensures stable yield. There are many definitions of stability, see e.g. McCann (2000), 
and we propose here to study the consequences of harvesting on stability through four different measures of conservation. The first two are impact measures; impact of harvesting on the population biomass and impact of harvesting on the population size structure. The second two measures have a natural link to the risk of extinction and will be referred to as risk measures. These are the resilience and the basic reproduction ratio (the recovery potential in case of the stage model).
We define the resilience as the reciprocal of the time needed for the population to recover from a perturbation, and we consider here both small and large perturbations.
The basic reproduction ratio/recovery potential describes the population's rate of increase from very low abundances, 
and thus could be construed as the likelihood of population rebound,
following a crash (e.g.~due to a large disturbance).

\subsubsection*{Measures of impact on biomass and size structure}

Let $J_{\text{}}^* = J_{\text{}}^*(F_{\text{J}},F_{\text{A}})$ and $A_{\text{}}^* = A_{\text{}}^*(F_{\text{J}},F_{\text{A}})$ denote the
juvenile and adult biomass at equilibrium, respectively,
of the harvested population.
In case of the stage model $J^{*}$ and $A^{*}$ are given directly by the state variables at equilibrium.
For the age model, $J^{*}$ and $A^{*}$ are obtained through the formulas
\begin{align*}
J^{*} = \sum_{a = 0}^{a_{\text{mature}} - 1} N^*_{a} s_{a} \quad \text{and} \quad
A^{*} = \sum_{a = a_{\text{mature}}}^{a_{\text{max}}} N^*_{a} s_{a},
\end{align*}
where $N^*_{a}$ denotes the number of individuals of age $a$ at equilibrium.
Furthermore, let $J_{\text{u}}^*$ and $A_{\text{u}}^*$ denote the juvenile and adult biomass at equilibrium
in the absence of harvesting, i.e. $J_{\text{u}}^* = J_{\text{}}^*(0,0)$ and $A_{\text{u}}^* = A_{\text{}}^*(0,0)$. 
Moreover, let $B_{\text{}}^* = J_{\text{}}^* + A_{\text{}}^*$
and $B_{\text{u}}^* = J_{\text{u}}^* + A_{\text{u}}^*$.
We measure \textit{impact on biomass} of harvesting through the expression
\begin{align*}
\text{Impact on biomass}\, =\, 1 - \frac{B^*_{\text{}}}{B^*_{\text{u}}}.
\end{align*}
%
%
Similarly, we consider \textit{impact on size-structure} through the expression
\begin{align*}
\text{Impact on size-structure}\,
=\, \frac{J_{\text{}}^*}{J_{\text{}}^* + A_{\text{}}^*} \left[ \frac{J_{\text{u}}^*}{J_{\text{u}}^* + A_{\text{u}}^*}\right]^{-1} - 1,
\end{align*}
which equals the relative change in the fraction of juvenile biomass following harvesting.
If the impact on size-structure is positive (negative),
then harvesting has increased (decreased) the fraction of juveniles in the population.\\




\subsubsection*{Resilience, basic reproduction ratio and recovery potential as risk measures}

Resilience as a risk measure is increasingly used in ecology
(Pimm and Lawton 1977; Loreau and Behera 1999; Petchey et al. 2002; Montoya et al. 2006; Loeuille 2010; Valdovinos et al. 2010).
Resilience is now also increasingly discussed in a fishery management context
(Hsieh et al. 2006; Law et al. 2012; Fung et al. 2013). 
The higher the resilience, the smaller the risk of extinction due to random drift.

We consider \textit{resilience} of the population
by measuring the reciprocal of the time needed for the population to recover the positive equilibrium given a random perturbation.
We do this by considering a large number of initial conditions.
From each initial condition, we measure the time until the population (and also the resource in case of the stage model)
returns to a small neighborhood of the equilibrium.
The average value of this return time over the number of trials are then used to quantify the resilience:
\begin{align*}
\text{Resilience} \,=\, \frac{1}{\text{Average value of the return times}}.
\end{align*}
Our resilience measure estimates the population's expected rate of return, given a random perturbation.
In contrast to many other studies on resilience that assess resilience based on eigenvalues of the Jacobian matrix,
our approach is not limited to the immediate neighborhood of the equilibrium,
but can also tackle large disturbances,
a point we will return to in the discussion section.
The precise procedure by which we determine the resilience is described in the Appendix
where we also present an alternative resilience measure,
estimating the population's probability to return within a time limit,
and reproduce some of our results using different magnitudes of disturbances.


We also consider the \textit{basic reproduction ratio} as a risk measure,
which represents the average number of offspring produced over the lifetime of an individual in the absence of density-dependent competition, i.e., when the population abundance is very low.
For the age model, we derive the following expression for the basic reproduction ratio as functions of the
harvesting rates $F_{\text{J}}$ and $F_{\text{A}}$:
\begin{align}\label{eq:basic-reprod-measure}
%
\text{Basic reproduction ratio}\,
=\, \frac{\left(1 - F_{\text{J}}\right)^{a_{\text{mature}}} \times \sum_{a = a_{\text{mature}} + 1}^{a_{\text{max}}} s_a\, S^{a} \left( 1 - F_{\text{A}} \right)^{a - a_{\text{mature}}}}
{ \left( 1 - \frac{{{h}}-0.2}{0.8 {{h}}}\right) \times \sum_{a = a_{\text{mature}} + 1}^{a_{\text{max}}} s_a \,S^{a} }.
\end{align}
%
%
%
A basic reproduction ratio larger than one ensures that the biomass of an initially small population increases on average, while a basic reproduction ratio less than one implies that the population will eventually become extinct.
The derivation of expression \eqref{eq:basic-reprod-measure} can be found in the Appendix.


%
In case of the stage model we use the \textit{recovery potential} introduced in \cite{MLBB13},
%
\begin{align*}
\text{Recovery potential}\, =\,  \frac{w_{\text{A}}(R_{\max})}{M + F_{\text{A}}}
\times \frac{v(w_{\text{J}}(R_{\max}))}{v(w_{\text{J}}(R_{\max}))-w_{\text{J}}(R_{\max}) + M + F_{\text{J}}}.
\end{align*}
The recovery potential is the generational net biomass production (per unit body mass) in a pristine environment
(free from density-dependent mortality)
and is therefore closely related to the basic reproduction ratio.
Similar as for the basic reproduction ratio, a recovery potential larger than one ensures that the biomass of an initially small population increases on average, while a recovery potential less than one implies that the population will eventually become extinct.
The basic reproduction ratio, as well as the recovery potential,
are directly linked to the probability of surviving a period of low population abundance during which random drift caused by demographic stochasticity can lead to extinction.
We further discuss this fact, as well as giving overall motivations of our choices of conservation measures,
in the discussion section.


\subsubsection*{Maximum sustainable yield and trade-off through Pareto efficiency}

Recalling that $J^{*}$ and $A^{*}$ denote the juvenile and adult biomass at equilibrium,
for any given harvesting rates $F_{\text{J}} \geq 0$, $F_{\text{A}} \geq 0$,
the yield objective function is given by
\begin{align}
\text{Yield}\, =\, F_{\text{J}}J^{*}  +  F_{\text{A}}A^{*}.
\end{align}
Moreover, the maximum sustainable yield (MSY) is obtained by taking the maximum of
the yield objective function across all harvesting strategies $(F_{\text{J}}, F_{\text{A}})$.
In addition to the yield function we are,
in case of both the age model and the stage model,
armed with four measures of conservation as functions of the harvesting rates $(F_{\text{J}}, F_{\text{A}})$.
Using these objective functions
we can calculate both the yield and the conservation for given harvesting strategies,
see Fig. \ref{fig:explaining_objectives} in the Results section.
%

To determine the trade-off between the two objectives yield and conservation,
we plot the yield as a function of each conservation measure in the results section
and apply the economic concept of Pareto efficiency to evaluate different harvesting strategies.
A harvesting strategy is Pareto efficient if it cannot be improved upon without trading off one of the considered objectives against the other, 
see e.g. Karpagam (1999, page 11).
The Pareto front is the set of all Pareto efficient harvesting strategies.
Hence, managers can restrict the choice of harvesting strategy to this set,
rather than considering the full range of possible harvesting strategies.
The closer a strategy is to the Pareto front, the more efficient it is.





\section*{Results}

\begin{figure}
\centering
\includegraphics[height=6.5cm,width = 7.5cm]{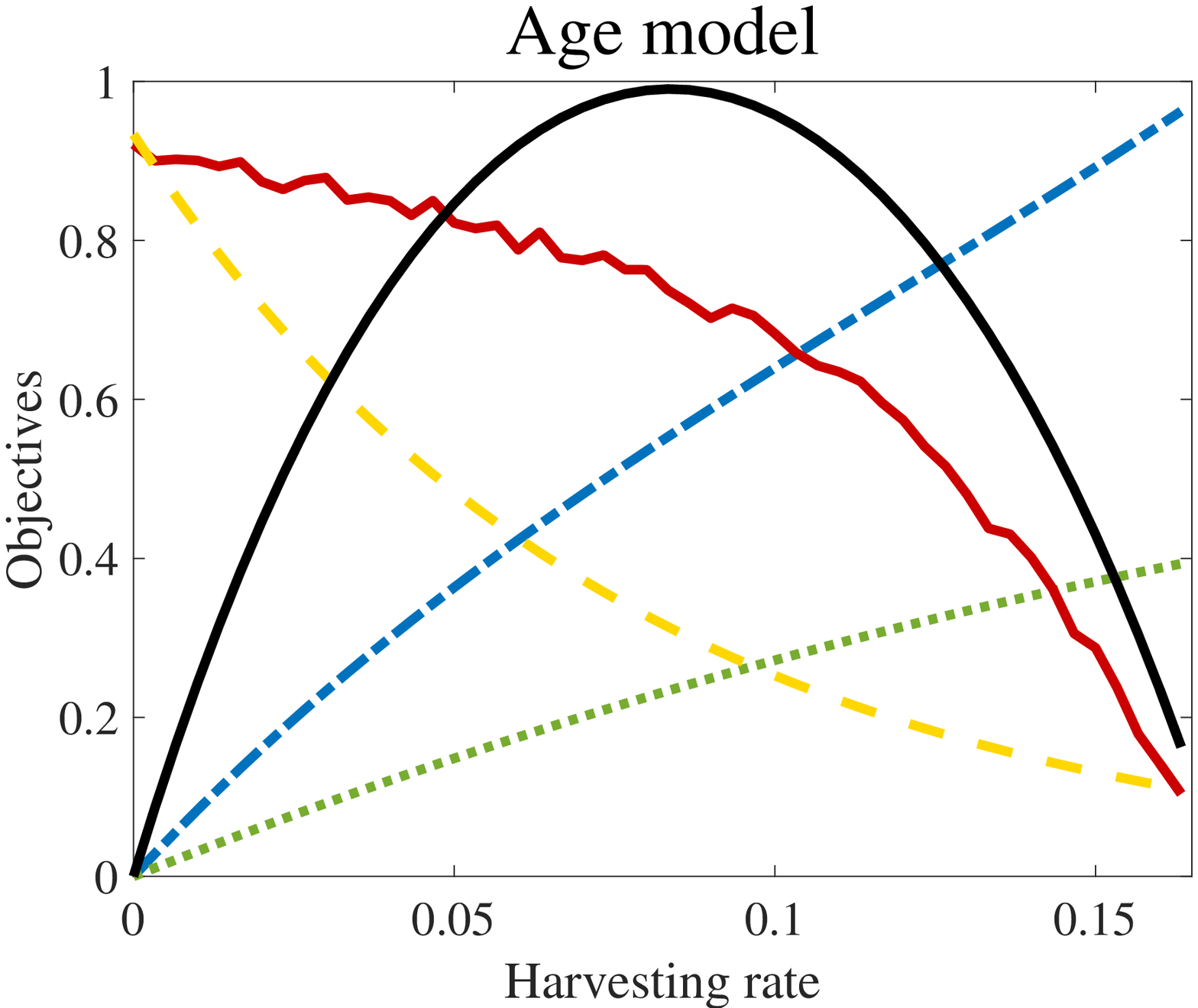}\hspace{0.5cm}
\includegraphics[height=6.5cm,width = 7.5cm]{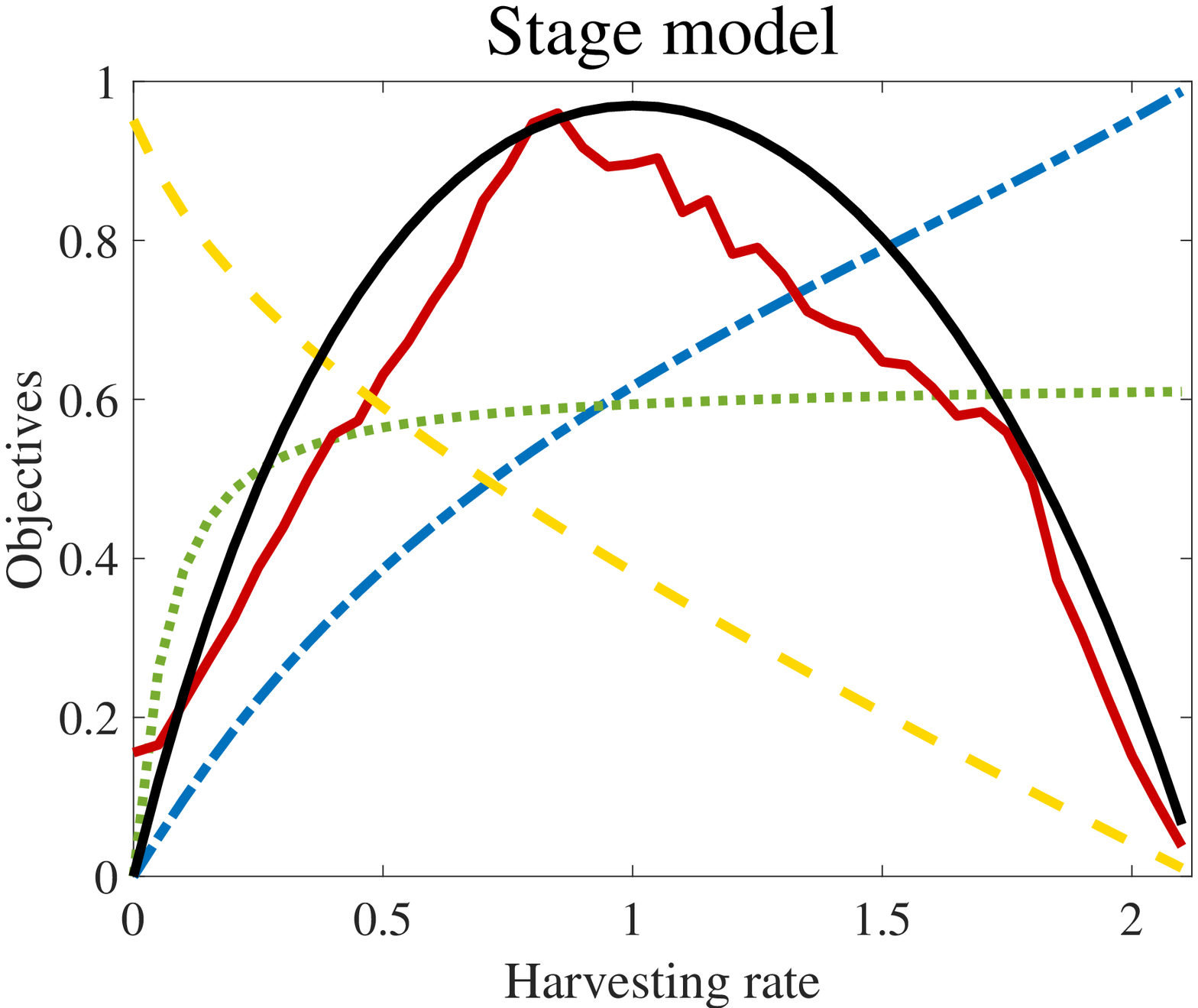}\hspace{0.5cm}
\caption{
The four measures of conservation and yield as functions of the harvesting rates considering equal harvesting  $(F_{\text{J}} = F_{\text{A}})$ in case of
the age model and the stage model.
Yield (black, solid),
impact on biomass (blue, dash-dot),
impact on size-structure (green, dotted),
resilience (red, solid) and
basic reproduction ratio (recovery potential) (yellow, dashed).
To visualize all objectives in the same plot graphs show (left) $14.5 \times$Yield, $50 \times$Resilience, $0.1 \times$Basic reproduction ratio
and (right) $3.3 \times$Yield, $5 \times$Resilience, $0.13 \times \log(\text{Recovery potential})$.
}
\label{fig:explaining_objectives}
\end{figure}

Figure \ref{fig:explaining_objectives} shows how the four measures of conservation and the yield changes with harvesting intensity for equal harvesting rates of juveniles and adults (henceforth \emph{equal harvesting}),
i.e.~$F_{\text{J}} = F_{\text{A}}$.
As harvesting pressure increases, the yield first increases 
after which it decreases as the population becomes ``overexploited".
The impact on biomass and the impact on size-structure increase with harvesting pressure, while the
basic reproduction ratio and the recovery potential decrease.
The resilience decreases with harvest pressure in case of the age model,
but first increases to a maximum and then decreases in case of the stage model.
Note that due to the different nature of the age model and the stage model,
the values of the harvesting rates in the two models may not be immediately compared.

We are now ready to present the trade-offs between yield and the four measures of conservation.
Figure \ref{fig:H1} represents results from the age model,
while Fig. \ref{fig:Roos1} gives the corresponding results for the stage model.

\subsection*{Pretty good yield allows large conservation benefits}

\begin{figure}
\centering
\includegraphics[height=15cm,width = 17cm]{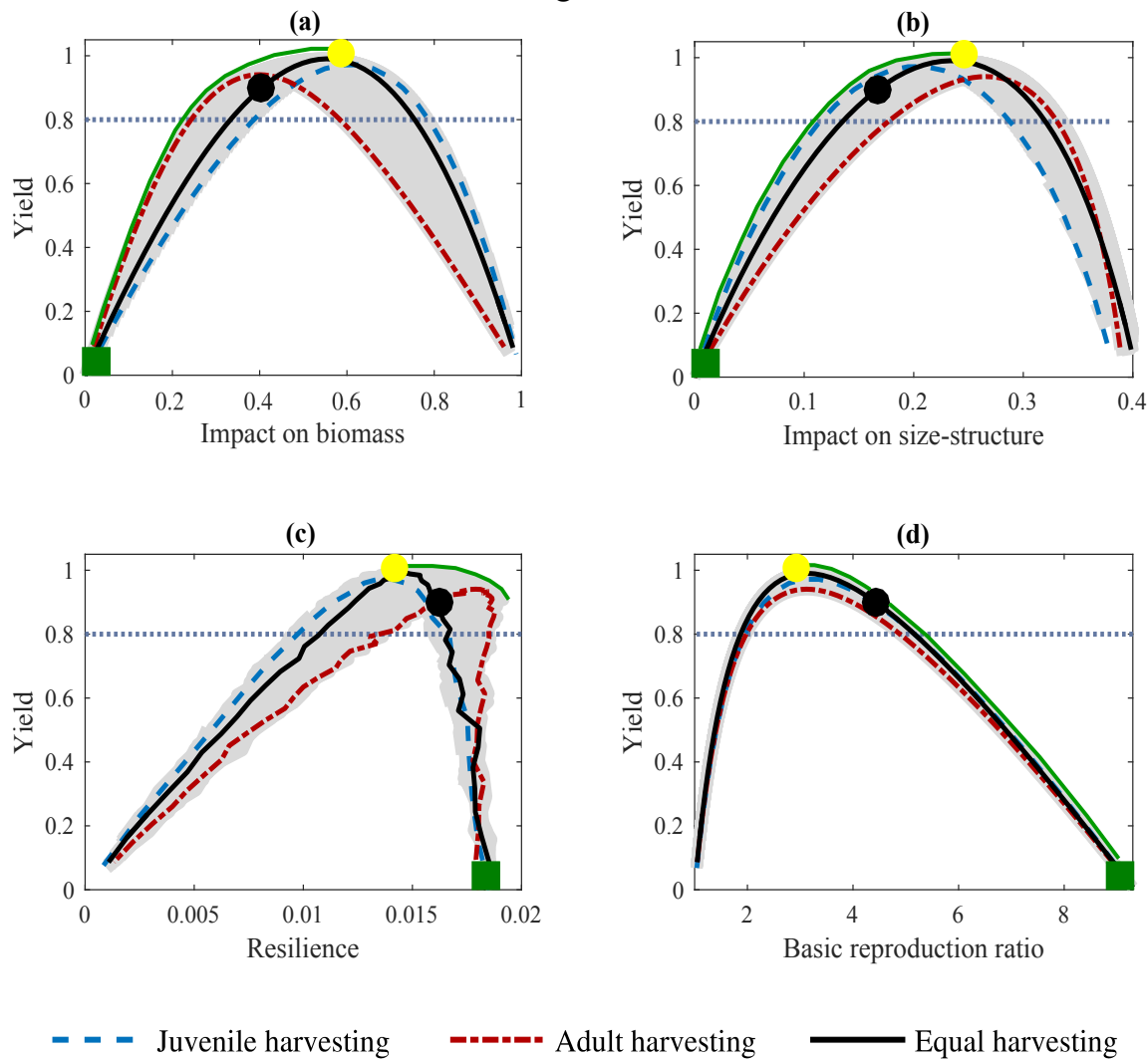}
\caption{
Trade-offs between yield and the four conservation measures in the age model.
The gray regions show ``all possible" combinations that can be realized when varying the harvesting rates on juveniles and adults. 
The solid green curves represent the Pareto front,
while the dotted grey lines give the border for PGY, i.e. $80\%$ of MSY.
The yellow dots represent MSY and the green squares give the unfished state.
We observe that within the range of PGY, equal harvesting performs well with respect to all measures.
The black dots represent a suggested harvesting strategy,
within the range of PGY, produced by $F_{\text{A}} = F_{\text{J}} = 0.06$.
Parameter values are as in \eqref{eq:param_age} and yield normalization is as in Figure \ref{fig:explaining_objectives}.
}
\label{fig:H1}
\end{figure}

\begin{figure}
\centering
\includegraphics[height=15cm,width = 17cm]{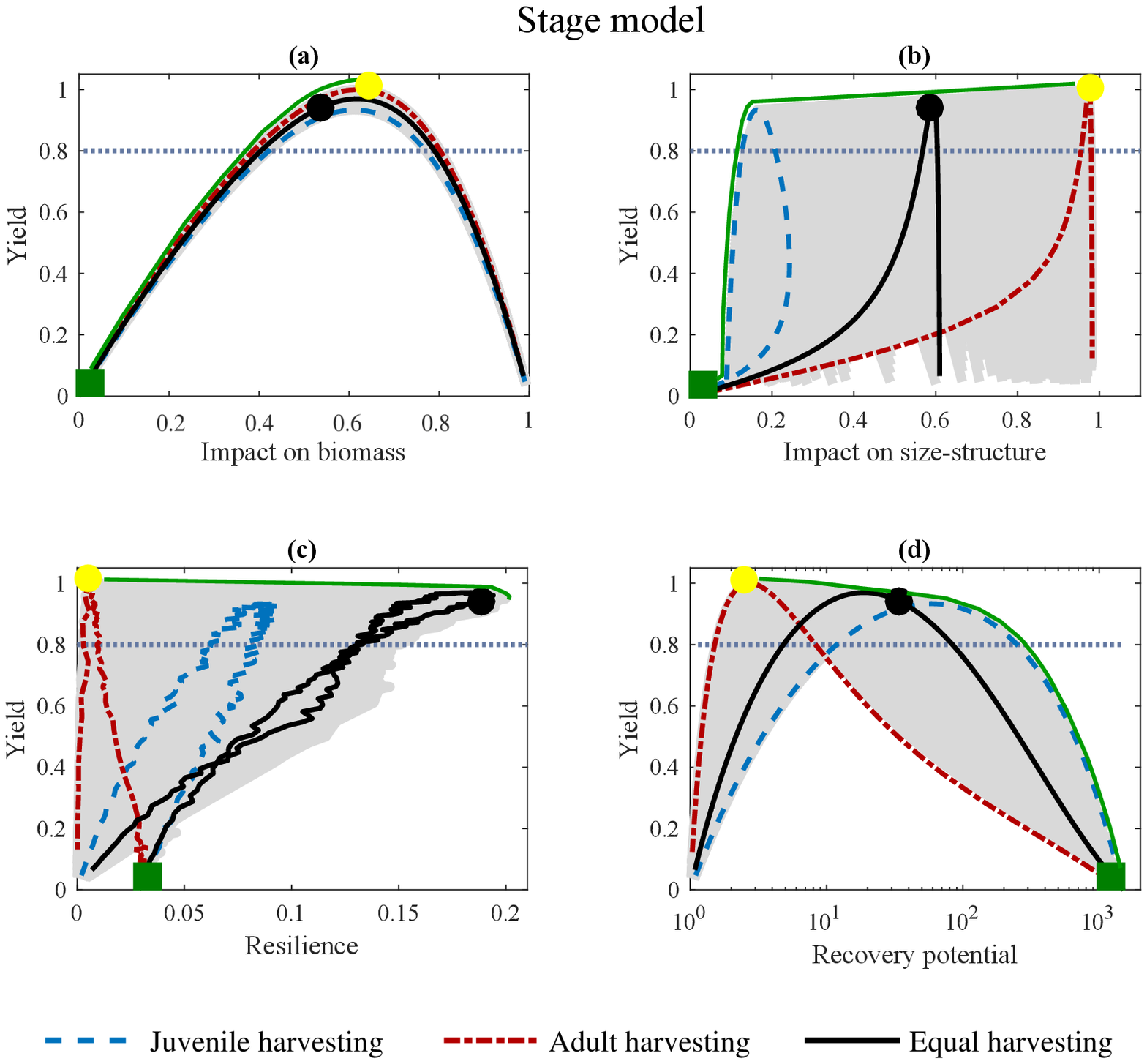}
\caption{
Trade-offs between yield and the four conservation measures in the stage model.
The gray region, curves, dots, and squares are as in Fig. \ref{fig:H1}.
We observe that within the range of PGY, equal harvesting performs well with respect to all measures.
The black dots represent a suggested harvesting strategy,
within the range of PGY, produced by $F_{\text{A}} = F_{\text{J}} = 0.8$.
Parameter values are as in \eqref{eq:param_stage} and
yield is normalized as in Fig. \ref{fig:explaining_objectives}.
}
\label{fig:Roos1}
\end{figure}

Focusing on the age model, Figs. \ref{fig:H1} (b)-(d) show that
the basic reproduction ratio is relatively low and the impact on size structure is also relatively large at MSY,
while the resilience is relatively high at MSY.
Focusing on the stage model, Figs. \ref{fig:Roos1} (c) and (d) show slightly different results; harvesting for MSY
(which is obtained by harvesting only adults)
gives a resilience and a recovery potential that is only a tiny fraction of the unexploited state and is close to the boundary of extinction.
Hence, harvesting for MSY may substantially increase the risk of stock collapse.
Fig. \ref{fig:Roos1} (b) shows also that the impact on size structure is at a maximum at MSY.

Common for both models and all four measures is,
see Figs. \ref{fig:H1} (a)-(d) and Figs. \ref{fig:Roos1} (a)-(d),
that by stepping back in yield by $20\%$ into the range of PGY,
we can find harvesting strategies with nearly half the impacts on biomass and half the impact on size structure,
and also with nearly twice the basic reproduction ratio (age model) and a much higher recovery potential (stage model).
Resilience can also be improved in case of both models,
though the difference in resilience is most
impressive for the stage model, see Figs. \ref{fig:Roos1} (c).
Hence, both the age model and stage model give the result that PGY allows for large conservation benefits.
Varying parameter values show that this conclusion is robust in both models (see Appendix).

However, stepping back in yield into the range of PGY
does not automatically ensure conservation in terms of any of the measures we consider.
To exclude the non-optimal harvesting strategies and to find the best ones within the range of PGY,
we apply the economic concept of Pareto efficiency, as introduced in the previous section.
Following the Pareto front
(the set of all Pareto efficient harvesting strategies shown as the green curves in
Figs. \ref{fig:H1} and \ref{fig:Roos1})
reveals these preferable harvesting strategies.
In the following, we will discuss simple harvesting strategies which are relatively close to the Pareto fronts in all cases.

\subsection*{Equal harvesting rates on juveniles and adults is often a good strategy}

Figs. \ref{fig:H1} and \ref{fig:Roos1} show that equal harvesting $(F_\text{J} = F_\text{A})$,
performs well with respect to both models and all four measures of conservation.
In particular, in the range of PGY, the black curves come rather close to the Pareto front in all subfigures
(especially in a neighborhood of the black dots).
Therefore, we can harvest juveniles and adults at equal rates,
which should be strategies that are rather easy to implement,
without losing too much yield or conservation.
The black dots in Figs. \ref{fig:H1} and \ref{fig:Roos1} show one such strategy.
Indeed, harvesting only adults is costly on some aspects, particulary in terms of resilience (stage model)
and the impact on size structure as well as basic reproduction/recovery potential (both models).
Harvesting only juveniles is costly in terms of resilience (both models) and in terms of impact on biomass (age model).

Varying parameter values show that this conclusion is very robust in the stage model,
where it seems to remain in the wide ranges
%
%
%
$
M \in [0, 0.5], \quad z \in [0.0001, 0.2], \quad  \sigma I_{\text{max}} \in [3,100], \quad q \in [0.6, 2], \quad R_{\text{max}} \in [0.5,100].
$
%
We refer the reader to the Appendix for substantial investigations
(both numerical and analytical) of robustness with respect to variations of
parameter values. Additional trade-off curves, as those presented in Fig. \ref{fig:Roos1},
are given for six different parametrizations in Figs. A 8, A 9 and A 10.

\begin{figure}
\centering
\includegraphics[height=12cm,width = 14cm]{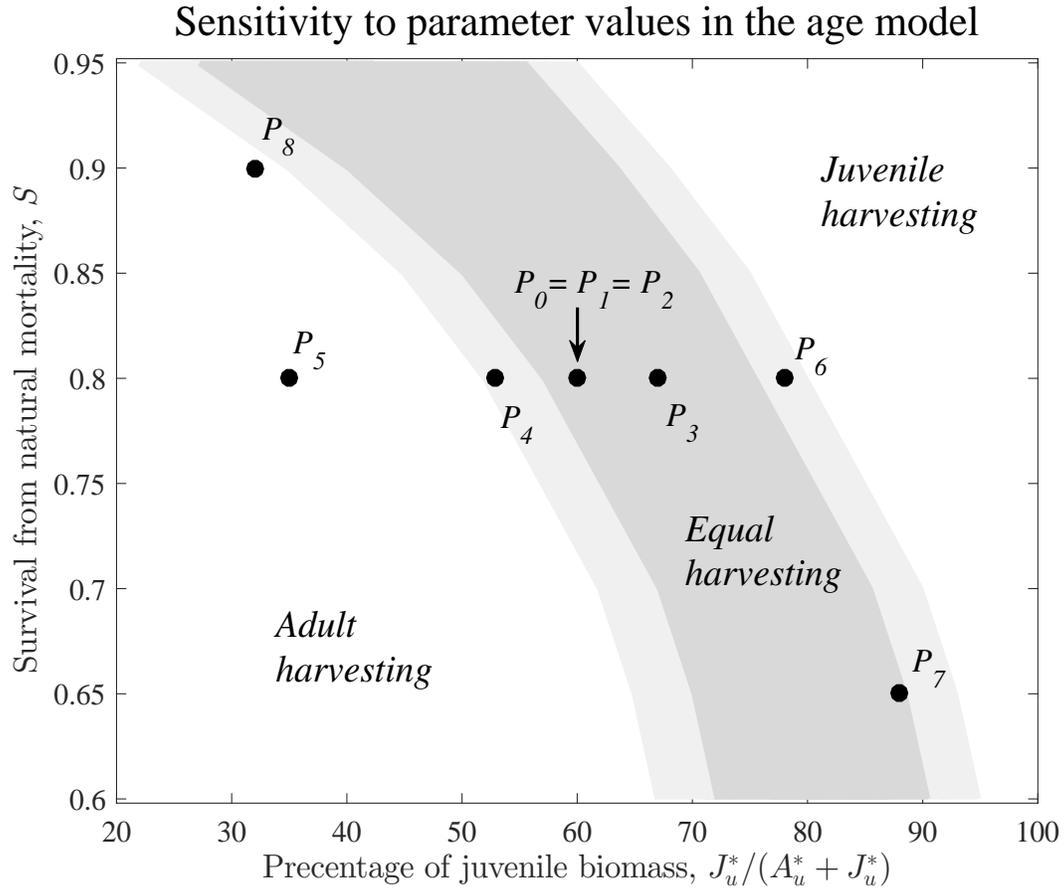}
\caption{
The harvesting strategy suggested by the results of the age model depends on $S$ and the fraction of juveniles in the unharvested population.
In the dark-grey region, equal harvesting is suggested by the age model,
while adult harvesting is better for low fraction of juveniles
and juvenile harvesting is to recommend when the fraction of juveniles is high. 
The light-grey regions refer to borderline cases.
Juvenile or adult harvesting is suggested only if such strategies are the most Pareto-efficient once,
within the range of PGY,
with respect to all four conservation measures.
Point $P_0$ corresponds to the default parametrization,
while points $P_1 - P_8$ correspond to parametrizations considered in the Appendix.
}
\label{fig:Sensitivity_H}
\end{figure}

However, equal harvesting is often but not always suggested by the age model.
Here, the efficient strategies seem to depend on the fraction of juveniles at the unharvested equilibrium,
$J^*_u/(A^*_u + J^*_u)$,
as well as on the survival from natural mortality, $S$.
We proceed by investigating this dependence by comparing pure adult harvesting (henceforth \textit{adult harvesting}),
equal harvesting and pure juvenile harvesting (henceforth \textit{juvenile harvesting}) for a wide range of parameter values in the age model.
Figure \ref{fig:Sensitivity_H} gives an approximation of regions in which the age model
suggest adult harvesting,
equal harvesting
and juvenile harvesting.
Juvenile or adult harvesting is suggested only if such strategies are the most Pareto-efficient once,
within the range of PGY,
with respect to all four conservation measures.
The borders in Figure \ref{fig:Sensitivity_H} are approximations which are produced by examining
a large number of variants of Figure \ref{fig:H1}
for parameter values in the intervals
%
$
a_0 \in [-3,-0.2], \quad  K \in [0.1, 1], \quad  a_{\text{mature}} \in [3, 15], \quad {{h}} \in [0.3,0.9], \quad \sigma_u \in [0, 0.5].
$
%
Indeed, we varied each parameter at a time,
keeping the others at the default values given in \eqref{eq:param_age},
and tested at least 10 values in each interval.
Further parameter combinations have also been tested in order to refine the borders in Figure \ref{fig:Sensitivity_H}.
Points $P_0 - P_8$ in Fig. \ref{fig:Sensitivity_H} correspond to different parametrizations of the age model.
The default parametrization in \eqref{eq:param_age} gives $J^*_u/(A^*_u + J^*_u) \approx 0.6$, $S = 0.8$ and is marked with $P_0$.
In Appendix Figs A 2, A 4, A 5 and A 6 we present trade-off curves, 
similar to those in Fig. \ref{fig:H1}, for
different parametrizations corresponding to the remaining eight points $P_1 - P_8$.
The Appendix also contains motivations and explanations for the dependence shown in Fig. \ref{fig:Sensitivity_H}.

It turns out that if it is possible to obtain PGY for a wide range of harvesting strategies
(including adult, juvenile and equal harvesting),
then our conservation measures are in favor of equal harvesting.
When adult or juvenile harvesting performs better than equal harvesting,
it is usually because equal harvesting can not give a yield in the range of PGY.

\subsection*{The age model is more sensitive to variations in parameter values than the stage model}

Focusing on the age model we first note that for the parameter values 
used in Figs. \ref{fig:explaining_objectives} and \ref{fig:H1} we have a survival from natural mortality of $S = 0.8$ \citep{ref_S}
and the fraction of juveniles in an unharvested population, $J^*_u/(A^*_u + J^*_u) \approx 0.6$.
We conclude that in this case equal harvesting is a good strategy.
Varying the parameter values, it turns out that
an increase in the fraction of juveniles implies an increase in the yield obtained when harvesting only juveniles,
i.e. the blue curves will be lifted in Fig. \ref{fig:H1}.
Similarly, a decrease in the fraction of juveniles implies an increase in the yield obtained when harvesting only adults, i.e. the red curves will be lifted in Fig. \ref{fig:H1}.
This dependence, which is expected and natural, can be observed in both models,
but it is much stronger in the age model.
While Fig. \ref{fig:Sensitivity_H} gives an approximation of the borders between adult harvesting,
equal harvesting and juvenile harvesting,
a similar investigation on the stage model gives
a much larger region suggesting equal harvesting.
In particular, in the stage model,
the most Pareto-efficient strategies,
within the range of PGY,
seems to be dominated by equal harvesting
as long as $0.1 < J^*_u/(A^*_u + J^*_u) < 0.9$.
(For the parameter values used in Figs. \ref{fig:explaining_objectives} and \ref{fig:Roos1},
we have $J^*_u/(A^*_u + J^*_u) \approx 0.5$.)

In conclusion, for populations in the region where the age model suggests equal harvesting,
the age model and the stage model agree on similar results.
For populations outside of this region
the age model suggests adult harvesting,
or, for some rare parameter settings,
juvenile harvesting.


\subsection*{Impact on size structure and impact on biomass serve as warning signals}

\begin{figure}
\centering
\includegraphics[height=15cm,width = 17cm]{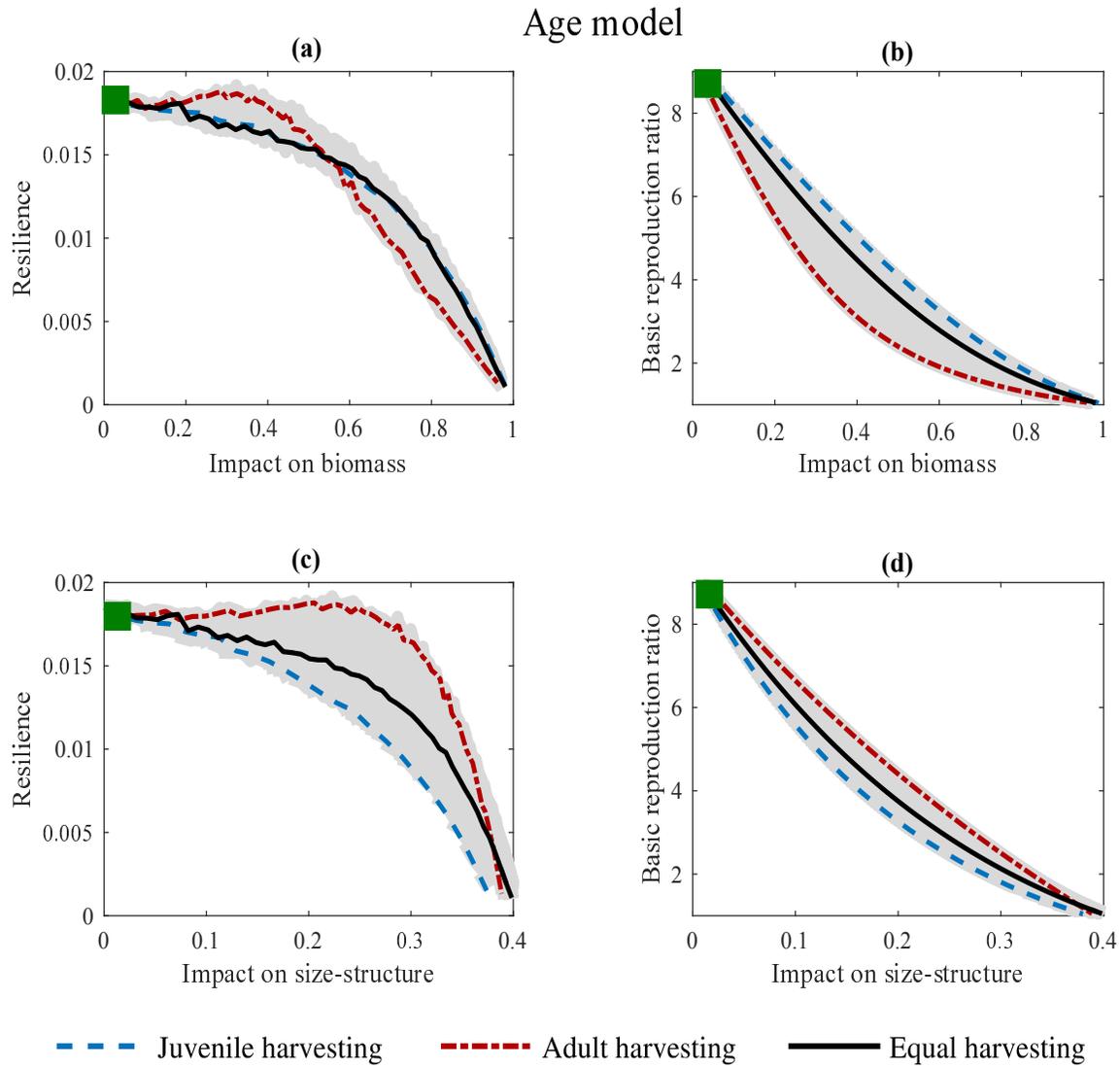}
\caption{
The relation between impact measures and risk measures for the age model.
We observe that a large impact on biomass implies a low basic reproduction ratio (recovery potential) and also a low resilience.
The same is true for impact on size structure.
Curves, green squares and parameters are as in Fig. \ref{fig:H1}.
}
\label{fig:H2}
\end{figure}

\begin{figure}
\centering
\includegraphics[height=15cm,width = 17cm]{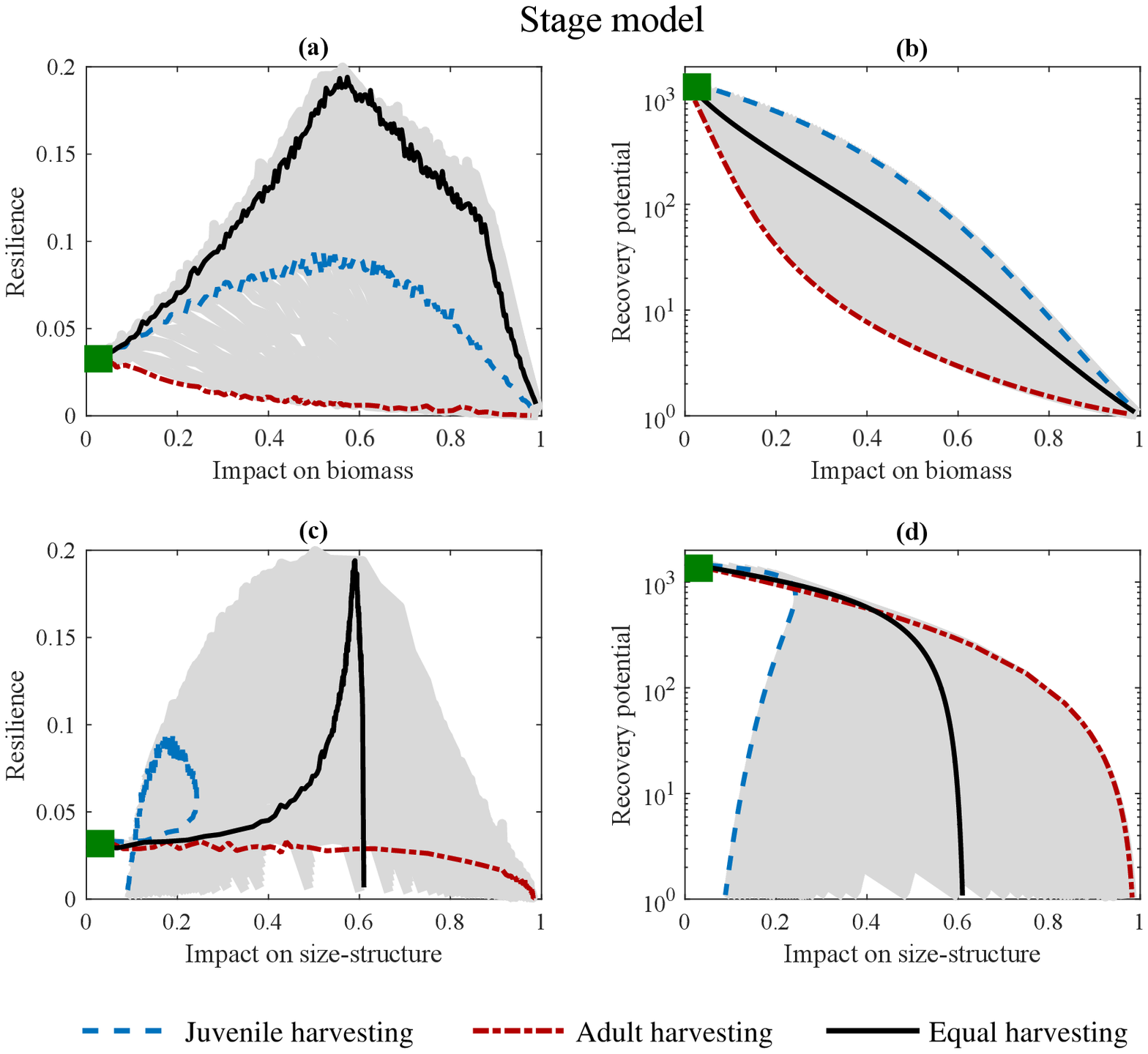}
\caption{
The relation between impact measures and risk measures for the stage model.
We observe that a large impact on biomass implies a low basic reproduction ratio (recovery potential) and also a low resilience.
The same is true for impact on size structure.
Curves, green squares and parameters are as in Fig. \ref{fig:Roos1}.
}
\label{fig:Roos2}
\end{figure}

As neither the resilience nor the basic reproduction ratio (recovery potential) can be directly measured in the field,
it is important to identify reliable proxies for conservation management that can be measured in field surveys.
Figs \ref{fig:H2} and \ref{fig:Roos2} show that a harvesting strategy with a high impact on population size structure,
or a high impact on biomass,
implies a low basic reproduction ratio (recovery potential) and a low resilience and hence a high risk of collapse.
%
%
Indeed, we find that resilience and basic reproduction ratio (recovery potential)
are systematically negatively correlated with impacts on biomass and size structure,
so that these later quantities,
which should be relatively easy to measure in field surveys,
can provide integrative signals to detect possible collapses.

\section*{Discussion}

We have investigated how well stage-dependent harvesting strategies that
qualify for pretty good yield (PGY) can account for conservation as a second objective.
To increase the chances that our results apply to a broad range of populations,
we have studied two established population models and reported conclusions that are common to both.
We have also investigated a wide range of parameter values for both models.
To incorporate conservation as a second objective for our optimization procedure,
we have used four different measures of conservation applied to both the age model and the stage model.
First, this extended analysis allows us to conclude strong robustness of
the results when all measures agree for both models;
e.g., that there are large potential gains of using specific PGY harvesting strategies that often,
but not always, correspond to equal harvesting rates of juveniles and adults.
Second, we are able to discuss and compare both the two models as well as the four measures of conservation with each other.

\subsection*{Implications for management of harvested populations}

Our study supports the implementation of PGY.
Furthermore, our results support implementation through equal harvesting of juveniles and adults, 
in conjunction with regular surveys that aim to detect changes in population biomass and size structure.
Managers aiming to implement optimal regulations 
may want to parameterize the age and stage model (or other suitable population models) for the specific species in question.
A similar analysis as the one presented here can then be carried out and will give the specific harvesting strategy that maximizes conservation benefits, e.g. as described by the four conservation measures considered here, for a given target-value of sustainable yield.

Managers relying on other approaches may still be interested in assessing changes in the size structure of a population, as well as changes in biomass, as these are strongly linked to our risk measures and may thus serve as warning signals for an impending collapse.
In fisheries management, changes in size structure and biomass can be measured through trial fishing, reinforcing our conclusion that size-structure and biomass are appropriate proxies for the risk of collapse and possible extinction.

One should note that equal harvesting should be relatively easy to implement.
Indeed, in a single-species setting 
an equal harvesting strategy is implemented by setting the same harvesting rate over all sizes of individuals.
This joint rate should then be tuned against the smallest value giving the desired yield.
In a multi-species setting, the harvesting rate should still be the same over all ages/sizes within each species,
but the preferable rate may differ between species.
Naturally, the rate should be higher for species with a higher productivity,
and lower for species having lower productivity.
Related to this is the concept of balanced harvesting,
which has attracted considerable attention recently,
and aims to distribute ``a moderate mortality from fishing across the widest possible range of species, stocks, and sizes in an ecosystem, in proportion to their natural productivity, so that the relative size and species composition is maintained"  (Garcia et al., 2012).
While implementing balanced harvesting is difficult since such strategy may be selective within each species as well (since productivity may depend on age/size), see e.g. Reid et al. (2016),
our results show that size-structure can be preserved fairly well by implementing the simple strategy of equal harvesting.

We have optimized for yield and conservation, not for economic yield.
Therefore, depending on the market (price of small fish versus price of large fish),
managers may obtain different preferable harvesting strategies if aiming for
economic yield.

\subsection*{Why harvest juveniles? Differences and similarities between the age model and the stage model}

While harvesting individuals before they mature is a debated topic,
we have seen that both the age model and the stage model give arguments for equal harvesting rates of juveniles and adults.
Indeed, relying on the stage model this argument is robust with respect to variations in parameters values.
The age model is more sensitive and the suggested harvesting strategy varies between mainly equal harvesting and adult harvesting as a function of parameter values.
To understand these results we first recall (see Results) that if it is possible to obtain PGY for a wide range of harvesting strategies,
then our conservation measures are in favor of equal harvesting.
Therefore, we can focus the following discussion on when and why the models allow for such wide range of harvesting strategies.

By extensive numerical experiments we illustrate
this dependence for the age model in Figure \ref{fig:Sensitivity_H}.
Varying the parameters values in the age model,
it turns out that an increase in the fraction $J^*_u / (A^*_u + J^*_u)$ of juveniles
implies an increase in the yield obtained when harvesting only juveniles,
and that a decrease in the fraction of juveniles implies an increase in the yield obtained when harvesting only adults.
This is natural; when considering harvesting of a population consisting of mainly juveniles it is not possible to obtain a good yield by harvesting only adults.
From Fig. \ref{fig:Sensitivity_H} we also see that as the survival from natural mortality $S$
increases, the recommendation goes towards including more juveniles in the harvesting strategy.
A reason for this is that for small $S$ additional mortality through harvesting on young individuals implies that too few individuals survive and become adults and the population declines.

A corresponding parameter dependence, as described in Fig. \ref{fig:Sensitivity_H},
is much weaker in case of the stage model.
Indeed, as mentioned in the results section,
the stage model suggest equal harvesting for wide ranges of parameter values,
see the Appendix for more on this.
One reason for this difference in sensitivity of the recommended harvesting strategies between the two models,
with respect to parameter values,
is as follows.
The stage model explicitly models the resource $R(t)$ through the third equation in \eqref{eq:stage-model},
and reproduction, growth and maturity are assumed to be increasing functions of the resource.
Therefore,
removing adult or juvenile biomass through harvesting
results in more resource available for the remaining population,
which in turn increases biomass production through all three mechanisms reproduction, growth and maturity of juveniles.
This feedback implies that the dynamics of the stage model allows for wide ranges of efficient harvesting strategies.

On the other hand, the age model incorporates the Beverton-Holt spawner recruit curve in \eqref{eq:recruitment} for reproduction,
and, independent of the recruitment,
individuals are assumed to grow following the Bertalanffy growth curve in \eqref{eq:growth_curve_B}.
Growth and recruitment are thus assumed to be independent in the age model,
while they are dependent through the resource in the stage model.
This means that if some juveniles are removed by harvesting
it will not be in favor of the recruitment of newborns in case of the age model,
as it would be in case of the stage model.
Thus, it is more costly to harvest juveniles in the age model than in the stage model,
and therefore the age model more often suggests to leave small individuals,
let them grow, and catch them as adults.

In conclusion, the more extensive population-level feedbacks in the stage model makes the population productive for a wider range of harvesting strategies
than the age model does,
and the age model is more restrictive to juvenile harvesting than the stage model.
This explains why equal harvesting performs well through wider ranges of parameter values in the stage model,
than in the age model.


\subsection*{Importance of preserving population size structure}

Our advice is based on our finding that large impacts on size-structure generally implies a high risk of collapse as captured by our risk measures,
see Fig. \ref{fig:H2} and \ref{fig:Roos2}.
To reduce the impact of harvesting on population size structure,
it seems advisable to harvest juveniles as well as adults,
see Fig. \ref{fig:H1} and \ref{fig:Roos1}.
Thus, equal harvesting is more likely to preserve
the size structure than single-stage harvesting.
(A similar conclusion was reached by Jacobsen et al. 2014.) 
We have shown that large impacts on size structure typically indicate unfavorable readings of our risk measures.
Our work thus reinforces the conclusions from a large and growing number of studies
(considering both ecological and evolutionary aspects)
that argue for the importance of preserving the size structure of harvested populations.
These studies, which we discuss below, reinforce the importance of including impact on size structure
explicitly as an important conservation measure when discussing harvesting strategies.
In fact, not accounting for the impact on size structure explicitly in our analysis means that
we should find the recommended harvesting strategies from Figs. \ref{fig:H1} and \ref{fig:Roos1} (a), (c) and (d)
only, not including the Pareto efficiency in Figs.  \ref{fig:H1} (b) and \ref{fig:Roos1} (b).
This would result in a shift towards recommending adult harvesting,
especially in case of the age model.

From an ecological point of view,
our analysis that size structure largely impacts the structure and functioning of the system is in agreement with previous works.  Anderson et al. (2008)
show that populations with a larger fraction of juveniles have less stable population dynamics because of changes in demographic parameters, and, therefore,
suffers a larger risk of extinction.
(In this context, see also Wikstr\"om et al. 2012.)
Moreover, changes in the size-structure of the population affect the balance of intra- and interspecific competition \citep{LM13}.

From an evolutionary point of view, affecting the size-structure of a population can potentially induce changes in biological traits such as size-at-age and age-at-maturation.
One reason is that harvesting only large individuals creates a large mortality selective pressure so that only adults that reproduce early, and at small size, pass their genes \citep{ny_evol}.
A consequence may be evolution toward small individuals reproducing early,
which is generally not desirable from an ecological (low reproduction) nor from an economical (too small to be valuable) point of view
\citep{Grift03, OHLMBED04}.
Evolutionary changes may have large impact on economic profit and future management
\citep{evol1, evol2, evol3}, and may be difficult to reverse.
Studying the collapse of the northern cod (\textit{Gadus morhua}, Gadidae), it has been shown that, before government imposed a moratorium, the life history shifted towards maturation at earlier ages and at smaller sizes \citep{OHLMBED04}, suggesting fisheries-induced evolution of maturation patterns.
Moreover, a recent study provides experimental evidence for rapid evolution induced by changes in the population size-structure of a fished population \citep{WTCDRROC13}.
Significant genetic variation for production-related traits is also present in fished populations \citep{L00},
and Cameron et al. (2013)
experimentally demonstrate evolutionary changes, in response to harvesting juveniles or adults.
In Kuparinen and Meril\"a (2007)
the authors argue that we should stop targeting only large individuals to avoid evolutionary impact on fisheries.
See also
Garcia et al. (2012) and Law et al. (2012)
for more arguments for harvesting preserving population size structure.
In conclusion, we recommend that managers consider the impact on size structure and that they avoid large deviations from the size structure of a pristine, unharvested population.



\subsection*{Relations between the four measures of conservation}

We have considered conservation as a second objective, beyond yield,
in our optimization procedure.
To quantify conservation we have chosen two impact measures;
\emph{impact on biomass} and \emph{impact on size structure},
and two ``risk" measures; \emph{resilience} and \emph{basic reproduction ratio (recovery potential)}.
It is not obviously true, even though it is expected,
that the impact measures relate simply to the risk measures.
Therefore, we present Figs \ref{fig:H2} and \ref{fig:Roos2} which show that
a large impact on biomass, or size structure,
implies a low basic reproduction ratio (recovery potential) and also a low resilience.
From this fact we concluded that it is important to preserve size structure and biomass in order to preserve stability of the population,
and that impact on biomass and impact on size structure work as warning signals for a collapse.

From Figs. \ref{fig:H2} and \ref{fig:Roos2}
it is clear that the relations between the four measures of conservation are not simple.
This is clearest from Fig. \ref{fig:Roos2} (a) and (c) representing the stage model;
resilience can vastly differ from the other measures by increasing with harvest pressure for some harvesting strategies.
This phenomena, which bears resemblances to the paradox of enrichment \citep{Rosenzweig, Rip},
deserves attention since it is very strong in the stage model under equal harvesting,
but not present at all under adult harvesting.
This thus provides a substantial argument in favour of equal harvesting.
Using an alternative resilience measure,
we give further illustrations and explanations of this behaviour of the stage model in the Appendix,
see Figs. A 11 and A 12.
%
%


Comparing resilience simulations in Figs. \ref{fig:H1} (c) and \ref{fig:Roos1} (c)
we conclude that harvesting of only adults is among the best strategies in the age model,
while such strategy is among the worst using the stage model (considering resilience only).
The resilience in the stage model instead suggest equal harvesting rates.
This is because in the age model,
the population will return fast to the equilibrium when harvesting only adults, after a given perturbation.
In the stage model however, the population returns very slowly when harvesting only adults,
compared to the case of equal harvesting.
Hence, transient behavior, and therefore the resilience, behaves different in the models.




Concepts of stability are numerous in ecology (McCann 2000),
and how the different stability measures relate to one another is considered a timely and important question
(Donohue et al. 2013). 
Indeed, measures of stability in ecological systems is today an active research area, see e.g.
Neubert and Caswell (1997)
for alternatives to resilience,
Nimmo et al. (2015)
for discussions of resistance and resilience, and
Isbell et al. (2015) as well as Dunne et al. (2002)
for stability and its relations to biodiversity.
Importantly, our model suggests that our different conservation measures covary,
and may be usefully assessed through changes in biomass and size structure.

\subsection*{Motivations of our choice of conservation measures}

As our results may depend on our chosen conservation measures,
we consider here additional motivations and discussions concerning this topic. 
%
%
%
First, the measures impact on biomass and impact on size structure
are important to consider simply since they can be measured in reality.
Second, these measures are natural, simple and easy to interpret and
a large impact on biomass would certainly imply impact on the surrounding ecosystem.
Moreover, in the subsection \emph{Importance of preserving population size structure},
we further motivated the impact on size structure as a central measure,
based on the fact that population size structure is
important to preserve from both an ecological and evolutionary point of view. 



To motivate the basic reproduction ratio and the recovery potential as a risk measure,
we note that, as already mentioned in the methods section,
these measures are directly linked to the probability of surviving a period of low population abundance during which random drift caused by demographic stochasticity can lead to extinction.
To see this, consider a small population in a pristine environment in which all individuals are, for simplicity, assumed to be identical. In this case, the basic reproduction ratio (or the recovery potential) is simply the ratio between birth-rate $b$ and death-rate $d$, that is $\Theta = b/d$. As proved by Grimmett and Stirzaker (1992, page 272),
the probability of avoiding extinction through random drift is given by $1 - 1/\Theta$ if $\Theta > 1$, and zero if $\Theta \leq 1$. Hence, there is a direct link between the basic reproduction ratio (recovery potential) and
the probability of surviving a period of low population abundance;
a high basic reproduction ratio (recovery potential) ensures a high probability of surviving.
To justify the investigation of the effects of large disturbances that bring the population to small numbers,
so that density dependence can be ignored, placing individuals in a pristine environment,
we mention mass mortality events \citep{stability_2},
drastic climate variability such as heat waves, storms, and floods \citep{stability_3},
and heavily exploited ecosystems \citep{stability_4}. 



To motivate our choice of resilience as a risk measure we first note that,
when dealing with nonlinear models,
many works considers only local stability and local resilience measures
(based on eigenvalues of the Jacobian matrix).
However, such approach gives only information arbitrarily close to the equilibrium,
saying little about the basin of attraction
(the set of initial conditions attracted by the equilibrium).
If the equilibrium is locally stable but the basin of attraction is small,
then even a small perturbation can force the dynamics to jump to another attractor, 
having possibly dangerous behaviour.
A large and convex basin of attraction, with the equilibrium in the middle,
ensures that the population recovers a perturbation with a high probability.
Therefore, it is natural to use both the size and the shape of the basin of attraction as stability/risk measures  (Lundstr\"om and Aidanp\"a\"a 2007, Menck et al. 2013, Lundstr\"om 2018).
However, such ``nonlocal" stability measure does not deliver any information in our case because for both models the
equilibrium is the unique globally stable attractor 
(The basin of attraction is the whole positive space, see Methods).
We proceed toward a nonlocal resilience measure by invoking the next natural candidate,
the return time to equilibrium given a perturbation,
and define \textit{resilience} as the reciprocal of the expected time needed for the population to
retain the equilibrium (see Methods).
In contrary to many works on resilience,
our nonlocal approach can invoke effects from small as well as from large perturbations
which is in line with classical definitions of ecological resilience (Walker et al. 1969, Holling 1973).
Our resilience measure estimates the population's expected rate of return, given a random perturbation.
In the Appendix we present an alternative nonlocal resilience measure,
the \textit{basin-time resilience}, which
is based on the size of a subset of the basin of attraction from which trajectories return fast.
Basin-time resilience estimates the probability that the population recovers the equilibrium within a time limit.
Results from this measure strengthen our previous conclusions and are illustrated in Appendix Figs. A 11 and A 12.

In general, our approach to resilience is applicable to advanced nonlinear models
(with complicated dynamics involving multiple attractors)
as well as to simple linear models with one unique stable equilibrium.
We have chosen to impose perturbations by random sampling from a uniform distribution,
but any set of perturbations may be considered,
e.g. normally distributed from equilibrium or a deterministic choice.
One may also consider perturbations only in the juvenile-, adult- or the resource dimension.
Our resilience measures link the widely used local approach (analyzed through eigenvalues) with the nonlocal one that is usually considered relevant by ecologists (accounting for basins of attraction and large disturbances)
as we may consider 
various ranges of disturbances.
We refer the reader to Lundstr\"om (2018) for further
discussions and constructions of nonlocal stability and resilience measures as well as their relations to local measures.
For discussion on the use of local resilience in ecology and the fact that it can be difficult to assess from an
empirical point of view, see
Haegeman et al. (2016).

\subsection*{Topics for future research}

%
In both the age model and the stage model we considered individuals in only two stages, as juveniles or as adults.
We then considered harvesting strategies that allow for different mortality rates in these two stages. 
Using slightly generalized versions of the age model and the stage model,
many more possible harvesting strategies can be explored.
A natural first step is to consider harvesting on a size interval,
and this can later be extended to include several size intervals as well as more realistic descriptions of harvesting mortality as a function of size.
Classical works of Beverton and Holt (1957) and Holt (1958) consider separate harvesting rates
on each year/size class and show that given a fixed harvesting effort,
the yield is maximized if fish are caught at the size or age where cohort biomass is maximum.
Extending our modeling to allow for different harvesting rates
on each year/size class would allow for evaluating such result
with respect to our suggested measures of conservation.


We point out that even though the stage model is purely deterministic,
one strength of our work is to tackle, through the nonlocal resilience measure and the recovery potential,
how the population recover from small as well as large stochastic perturbations,
and how the population may survive demographic stochasticity at low density levels.
To account for more stochastic effects,
a possible direction would be to expand the modeling towards demographic and environmental stochasticity,
since the interplay between stochasticity of demographic parameters and deterministic nonlinearity is important \citep{letter}.
A study in this direction by Engen et al. (2018)
has found that adding environmental stochasticity may change predictions of which harvesting strategy
(adult, juvenile, or mixed harvesting) that gives the highest yield.
They show that even when a deterministic model gave the highest yield from adult or juvenile harvesting,
adding environmental stochasticity caused mixed harvesting to give the highest yield in many cases.
This indicates that our result ``equal harvesting is often a good strategy"
might be further strengthened when environmental stochasticity is given further account.

Another promising extension of the work presented here is to move beyond single-species management towards ecosystem-based management.
We believe in a trend from single-objective towards multi-objective approaches
(i.e. optimizing for yield and conservation, not only yield), strengthened by the present paper.
This trend may evolve towards multi-objective approaches using multi-species population models, that is, towards ecosystem management. Studies in this direction already exists, see e.g. White et al. (2012), Tromeur and Loeuille (2017)
and Jacobsen et al. (2017).
Our four measures of conservation can be extended to more general multi-species settings, and
the present method using Pareto frontiers to find sustainable harvesting strategies
can then be applied also in such general settings.
For example, harvesting on a set of $k$ species in a food web with respective harvesting rates
($F_{1}, F_{2}, ..., F_{k}$),
we can for any desired yield determine the harvesting strategy that offers the highest conservation benefits.
Hence, the methods presented here open a door for reconciling economic and conservation issues in ecosystem management and can be extended to more complex scenarios including for example management of multiple fisheries and maintaining species diversity.

A concept which has attracted considerable attention recently is balanced harvesting
(see e.g. Garcia et al., 2012), see also the beginning of Discussion.
Balanced harvesting strategies should preserve ecosystems' relative size and species composition,
and thus harvesting rates may need to be adjusted in proportion to the productivity of
individuals.
As productivity differs among species, and also within a single species,
a balanced harvesting strategy is probably selective and nontrivial to find and implement.
Law et al. (2015) argue that switching from size-at-entry regulations to balanced harvesting can increase both yield and conservation. 
Noting that equal harvesting is a more balanced strategy than adult harvesting, their result is in line with the present paper. 
As our approach can potentially be extended to a general multi-species settings
it can also be applied to evaluate general balanced harvesting strategies in the framework of advanced population models.



\vspace{20mm}

\begin{center}
\Large{\bf Appendix--supporting research for the manuscript Meeting yield and conservation objectives by harvesting of both juveniles and adults}
\end{center}
\normalsize

\renewcommand{\figurename}{Figure A}
\setcounter{figure}{0} 




\noindent
The appendix gives additional motivations, explanations and details on the main manuscript,
as well as additional numerical and analytical results which strengthen the main findings of the main text.
We begin by motivating parameter values as well as investigating robustness with respect to variations of parameter values in the age model, and proceed with a similar section for the stage model.
Next, we describe our resilience measure in detail and give some additional resilience investigations of the stage model using an alternative resilience measure.
We end by deriving the analytical expression for the basic reproduction ratio for the age model.

\section*{Motivations and variations of parameter values in the age model}

Concerning the parametrization of the age model,
we have used
\begin{align}\label{eq:Aparam_age}\tag{A1}
\mathcal{R}_0 = s_{\text{max}} = c &= 1, \quad a_{\text{max}} = 100, \quad K = 0.23, \quad a_0 = -2, \\ \nonumber
a_{\text{mature}} &= 8, \quad h = 0.7, \quad S = 0.8, \quad \sigma_u = 0,
\end{align}
as default values in the main text, and we have considered substantial variations from these particular values.
When varying parameter values in the age model,
we have seen that the preferable harvesting strategies depend on the parametrization in a way that can be well-explained by
Fig. 4 in the main text.
That is, the suggested harvesting strategy (when comparing adult harvesting, equal harvesting and juvenile harvesting) depends on the survival from natural mortality, $S$,
as well as on the fraction of juveniles in the unharvested case,
$J^*_{\text{u}}/(J^*_{\text{u}} + A^*_{\text{u}})$.
The results in Fig. 4 seem to be robust whenever parameters take on values in the wide intervals
\begin{align}\label{eq:Aparam_age_interval}\tag{A2}
 K \in [0.1, 1], \quad a_0 \in [-3,-0.2], \quad  a_{\text{mature}} \in [3, 15], \quad {{h}} \in [0.3,0.9], \quad \sigma_u \in [0, 0.5].
\end{align}
Indeed, we have varied each parameter at a time,
keeping the others at the values given in \eqref{eq:Aparam_age}
and tested at least 10 values in each interval.
Further parameter combinations in the above intervals have also been tested.

In the following we will give motivations for the parameter values in \eqref{eq:Aparam_age}
as well as for the considered intervals in \eqref{eq:Aparam_age_interval}.
We will also give some explanations of how and why our results depend, or not depend,
on the parameter values.
Finally, we present additional trade-off curves
(similar to those in Fig. 2 in the main text) but for parametrizations corresponding to
points $P_1 - P_8$ in Fig. 4 in the main text.
These curves are given in Figs. A \ref{fig:age_steepness},  A \ref{fig:age_a0},  A \ref{fig:age_amature} and A \ref{fig:age_S}.
%
Careful reading of these figures should convince the reader that our main findings
are robust with respect to variations of parameter values.
We do not present additional versions of main text Fig. 5 (showing relation between measures).
However,
the correlation can be seen from the trade-off figures 
by first focusing on the MSY point and then follow curves, for different measures,
in the direction of either increasing or decreasing harvesting pressure.

\subsection*{Reduction of parameters}

We begin by motivating that without loss of generality we can choose $\mathcal{R}_0 = s_{\text{max}} = c = 1$,
as well as  $a_{\text{max}} = 100$,
and we will therefore not consider variations of these parameter values.
In particular, by careful investigation of the equations in the main text we see that 
if we divide $f_a = c s_a$ in eq. (1) by $c$,
eq. (2) by $s_{\text{max}}$ and
eq. (4) by $\mathcal{R}_0$, 
then $\mathcal{R}_0$, $s_{\text{max}}$, $c$, $s_a$, $E_t$, $R_t$ and $N_{a,t}$ 
can be considered as non-dimensional. 
We obtain non-dimensional equations, similar to the old once but with $\mathcal{R}_0 = s_{\text{max}} = c = 1$.
This shows why these parameters can be set to $1$ without loss of generality.
Indeed, the results for arbitrary $\mathcal{R}_0$ and $s_{\text{max}}$ can be obtained by multiplying the yield obtained for $\mathcal{R}_0 = s_{\text{max}} = 1$, given by main text eq. (11), by arbitrary values of these parameters.
The parameter $\mathcal{R}_0$ scales the number of individuals in the population, per unit volume, 
while $s_{\text{max}}$ scales the mass of each individual. 
The constant $c$ sets the number of eggs per individual, and 
the assumed Beverton-Holt recruitment relation \eqref{eq:Arecruitment}, 
giving the relation between egg production and offspring, 
saturates in relation to the number of eggs at the unharvested equilibrium. 
As our results are independent of the number of eggs in the lake, 
the value of the constant $c$ is unimportant.
This explains also that we can take $m_a = 1/2$ in place of $m_a = 1$ and consider half the population as adults;
that would only correspond to considering half the number of eggs per individual, i.e. $c = 1/2$. 

Finally, we will take the maximum age of an individual to be
so large that it only cut away a negligible amount of biomass, i.e., very old fish.
For our needs, $a_{\text{max}} = 100$ is enough.


\subsection*{Steepness $h$ in the Beverton-Holt recruitment, points $P_1$ and $P_2$}

We recall the assumed Beverton-Holt recruitment 
\begin{align}\label{eq:Arecruitment}\tag{A3}
R_{t+1} = \frac{E_{t}}{\alpha + \beta E_{t}} \text{exp} \left( u_t- \frac{\sigma_u^2}{2}\right),
\end{align}
in which $u_t$ are independent and normally distributed random variables with mean 0 and standard deviation $\sigma_u$.
The parameters $\alpha$ and $\beta$ are given by
\begin{align*}
\alpha = \frac{\mathcal{E}_0}{\mathcal{R}_0} \left( 1 - \frac{{{h}}-0.2}{0.8 {{h}}}\right), \qquad \beta = \frac{{{h}}-0.2}{0.8 {{h}} \mathcal{R}_0},
\end{align*}
where $\mathcal{E}_0$ and $\mathcal{R}_0$ are, respectively,
the average egg production and recruitment at equilibrium in the absence of harvesting mortality.
The parameter ${h}$ is the steepness which sets the sensitivity of recruitment with respect to egg production
and may take values between 0.2 and 1.
The steepness is defined as the ratio of recruitment when egg production equals $20\%$ of $\mathcal{E}_0$
to recruitment at $\mathcal{E}_0$ \citep{MD88, H10}.

The relation between $\mathcal{E}_0$ and $\mathcal{R}_0$ yields 
\begin{align}\label{eq:Arelation_E0_R0}\tag{A4}
\mathcal{E}_0 = \mathcal{R}_0 \sum_{a = 0}^{a_{\text{max}}}  m_a f_a S^{a}.
\end{align}
To derive this relation, put $R_t = \mathcal{R}_0$ and $\gamma_{a-1} = 0$ in main text eq. (4). 
It then follows that numbers of individuals at equilibrium, in the absence of harvesting, are
$$N_{0,t} = \mathcal{R}_0, \quad N_{1,t} = \mathcal{R}_0 S, \quad N_{2,t} = \mathcal{R}_0 S^2, \dots, 
N_{a_{\text{max}},t} = \mathcal{R}_0 S^{a_{\text{max}}}.
$$
By summing up these individuals total egg production (recall main text eq. (3)) 
we obtain relation \eqref{eq:Arelation_E0_R0}. 

We have already chosen $\mathcal{R}_0 = 1$ and the value of $\mathcal{E}_0$ follows from
$m_a$, $f_a$, $s_a$ via \eqref{eq:Arelation_E0_R0}.
Moreover, by numerical investigations we have convinced ourselves that there is little effect of varying $\sigma_u$.
Therefore, in the following we focus on the steepness parameter $h$.

Myers et al. (1999) reviewed the steepness of 244 stocks of fish
and found that steepness values varies mainly in the range from 0.3 and 0.9.
An intermediate value of 0.7 and slightly greater steepness values are common
(Myers et al. 1999, Table 1).
We have considered variations of steepness in the range $h \in [0.3,0.9]$,
and to clarify the dependence of steepness we present additional trade-off curves in Fig. A \ref{fig:age_steepness}
for $h = 0.5$ (point $P_1$ in Fig. 4) and $h = 0.9$ (point $P_2$ in Fig. 4).
It turns out that our main results are not sensitive to the steepness.
We can see in Fig. A \ref{fig:age_steepness} that $(i)$ our conclusion that PGY allows large conservation benefits
becomes more clear when steepness increases, and that $(ii)$
equal harvesting performs well rather independent of steepness.

\begin{figure}
\centering
\includegraphics[height=6.5cm,width = 7.5cm]{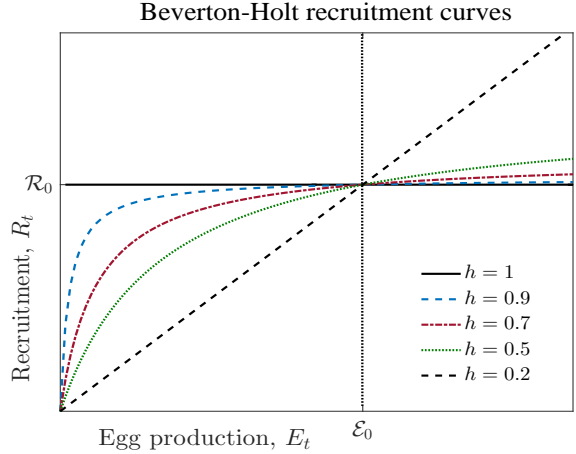}
\caption{Recruitment as functions of egg production following \eqref{eq:Arecruitment} for different values of the steepness parameter $h$.
When $h = 0.2$ then recruitment is proportional to egg production,
and when $h = 1$ then recruitment is constant, independent of egg production.
Remaining parameters are as in \eqref{eq:Aparam_age}.}
\label{fig:steepness-curve}
\end{figure}

\begin{figure}
\centering
\includegraphics[height=14.5cm,width = 16cm]{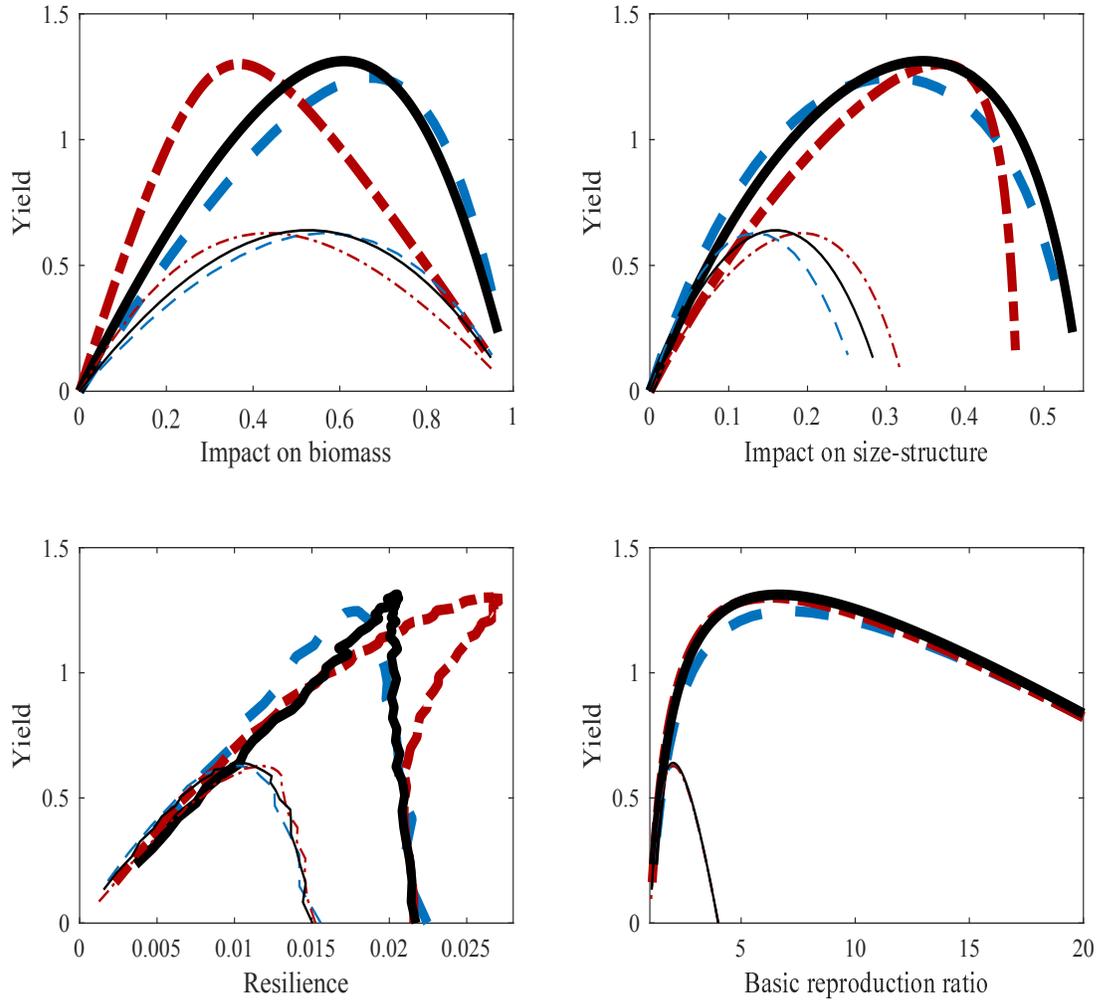}
\caption{Age model for $h = 0.5$ (thin curves) giving point $P_1$ in Fig. 4
and $h = 0.9$ (thick curves) giving point $P_2$.
Juvenile harvesting (blue, dashed), equal harvesting (black, solid), adult harvesting (red, dash-dot).
Remaining parameters are as in \eqref{eq:Aparam_age} and
yield is normalized as in Fig. 2 in the main text.
}
\label{fig:age_steepness}
\end{figure}

To understand why $(i)$ and $(ii)$ should hold in general,
and how our conservation measures depend on the steepness,
we proceed by noting the following.
If $h = 1$ then $\alpha = 0$, $\beta = 1/\mathcal{R}_0$ and $R_t = \mathcal{R}_0$ (constant recruitment).
If $h = 0.2$ then $\alpha = \mathcal{E}_0 / \mathcal{R}_0$, $\beta = 0$ and $R_t = E_t \cdot \mathcal{R}_0 / \mathcal{E}_0 $
(proportional recruitment), see Fig. A \ref{fig:steepness-curve}.
Thus, populations with high steepness values can be assigned heavy harvesting pressures,
implying small abundances, and still give high yield because the reproduction stays high also for small egg production.
To explain this, we note that at unharvested equilibrium we have $R_t = \mathcal{R}_0$ and $E_t = \mathcal{E}_0$,
i.e. where all curves intersect in Fig. A \ref{fig:steepness-curve}.
Imposing harvesting on the population will reduce the population biomass, and hence reduce the total egg production,
and we therefore move to the left of the point of intersection in Fig. A \ref{fig:steepness-curve}.
If steepness is high (close to 1), then recruitment is still close to $\mathcal{R}_0$,
but if steepness is low (close to 0.2), then recruitment decrease linearly.
From this reasoning we realize that high steepness and high harvesting pressure give a situation with
(relatively) high reproduction, high harvesting mortality and high yield.
Therefore, young individuals should be dominating,
and hence the impact on size structure and the impact on biomass will be high.
Moreover, the basic reproduction ratio as well as the resilience will be high.
These phenomena can be seen in Fig. A \ref{fig:age_steepness}.

From the above reasoning we also realize that $(i)$ should hold in general.
In fact, when steepness increases PGY can be obtained for wider ranges of harvesting strategies,
giving more possibilities to chose harvesting strategies (giving high yield) from.
See Hilborn (2010) for more on this.

To understand why $(ii)$ should hold in general,
we recall that at the unharvested equilibrium we have $R_t = \mathcal{R}_0$ and $E_t = \mathcal{E}_0$ independent
of the steepness $h$.
Hence, the fraction of juveniles at the unharvested equilibrium,
$J^*_{\text{u}}/(J^*_{\text{u}} + A^*_{\text{u}})$,
as well as the location of the parametrization in Fig. 4 in the main text,
are also independent of the steepness.
As the suggested harvesting strategy mainly depends on $J^*_{\text{u}}/(J^*_{\text{u}} + A^*_{\text{u}})$ and $S$,
it follows that it should be rather independent of the steepness $h$.

\subsection*{Parameters $a_0$ and $K$ in the von Bertalanffy growth, points $P_3$ and $P_4$}

\begin{figure}
\centering
\includegraphics[height=6.5cm,width = 7.5cm]{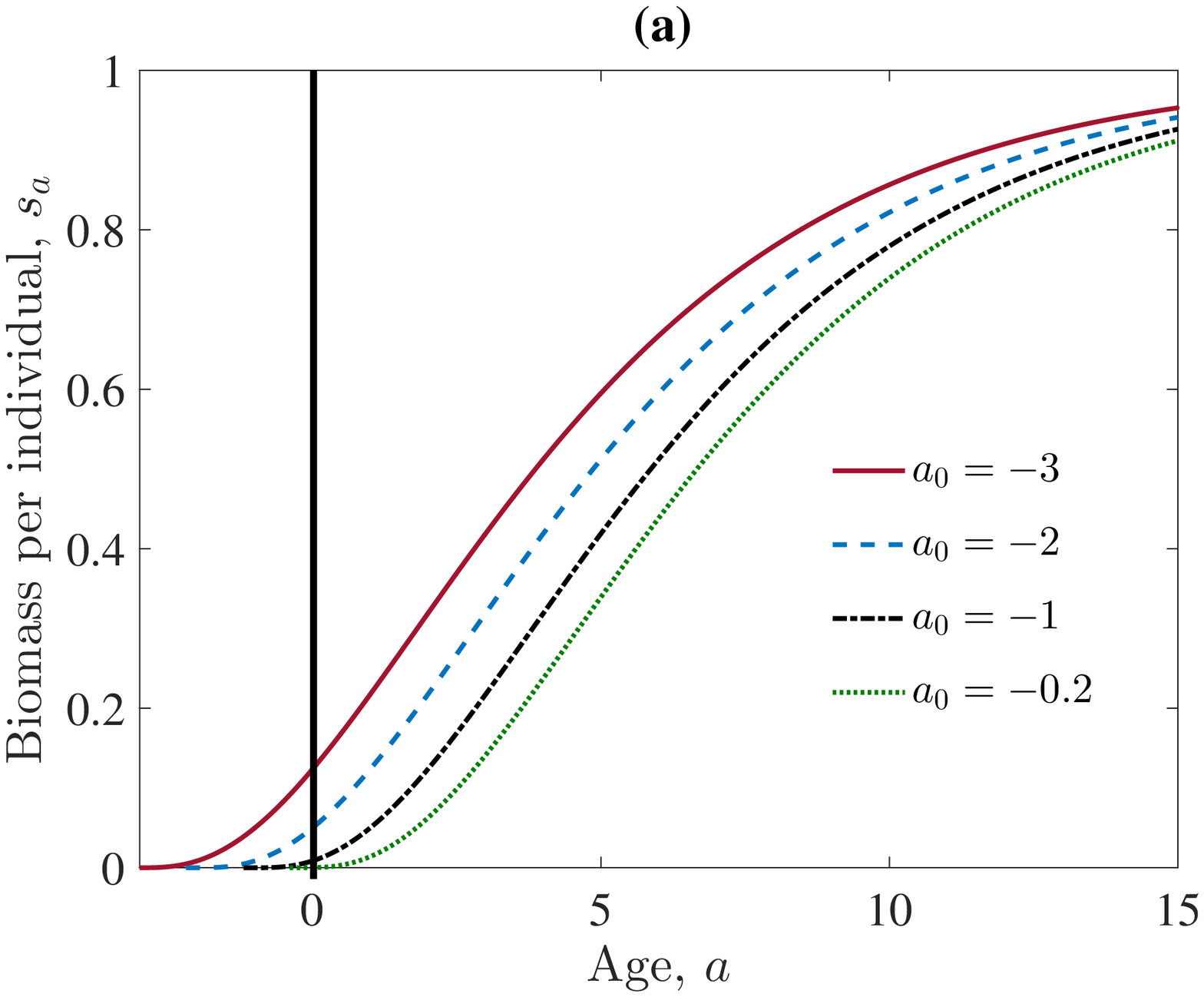}\hspace{0.1cm}
\includegraphics[height=6.5cm,width = 7.5cm]{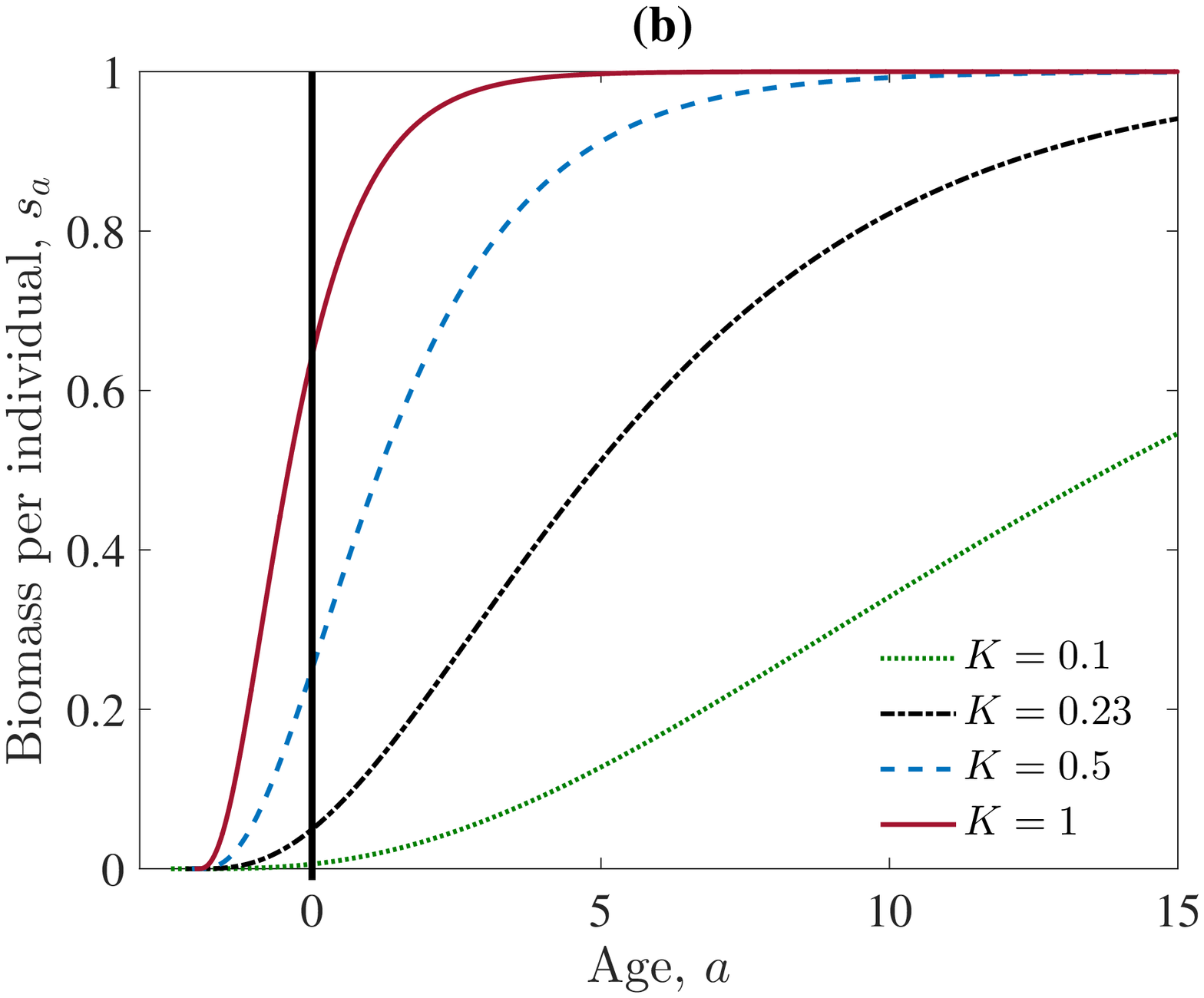}
\caption{Growth of an individual as functions of age following the von Bertalanffy growth curves \eqref{eq:Agrowth_curve_B} for different values of the parameters $a_0$ and $K$.
(a) variation of $a_0$, (b) variation of $K$. Remaining parameters are as in \eqref{eq:Aparam_age}.}
\label{fig:growth-curve}
\end{figure}

We have assumed von Bertalanffy (1957) 
growth to describe individual length as a function of age,
and that individual mass is proportional to the cube of individual length, i.e.
\begin{align}\label{eq:Agrowth_curve_B}\tag{A5}
s_a &=  s_{\text{max}} \left(1 - e^{-K(a - a_0)}\right)^{3},
\end{align}
where $s_a$ is the mass of an individual at age $a$,
$s_{\text{max}}$ is the asymptotic maximum body mass,
$K$ is a growth rate parameter and
$a_0$ is the hypothetical negative age at which the individual has zero length.
We have already motivated $s_{\text{max}} = 1$.
Punt et al. (1995, page 290) studying the albacore (\textit{Thunnus alalunga}, Scombridae)
motivate us to take $a_0 = -1$ and $K = 0.23$.
However, the value of $a_0$ may be decreased to $-2$ as well since
the age model is discrete and calculates the size of all individuals in the beginning of the year, but
harvesting and egg production naturally occur continuously during the year.
We have considered variations in the ranges
$a_0 \in [-3,-0.2]$ and $K \in [0.1, 1]$.
Figure A \ref{fig:growth-curve} shows the corresponding growth curves for some values of $a_0$ and $K$.

\begin{figure}
\centering
\includegraphics[height=14.5cm,width = 16cm]{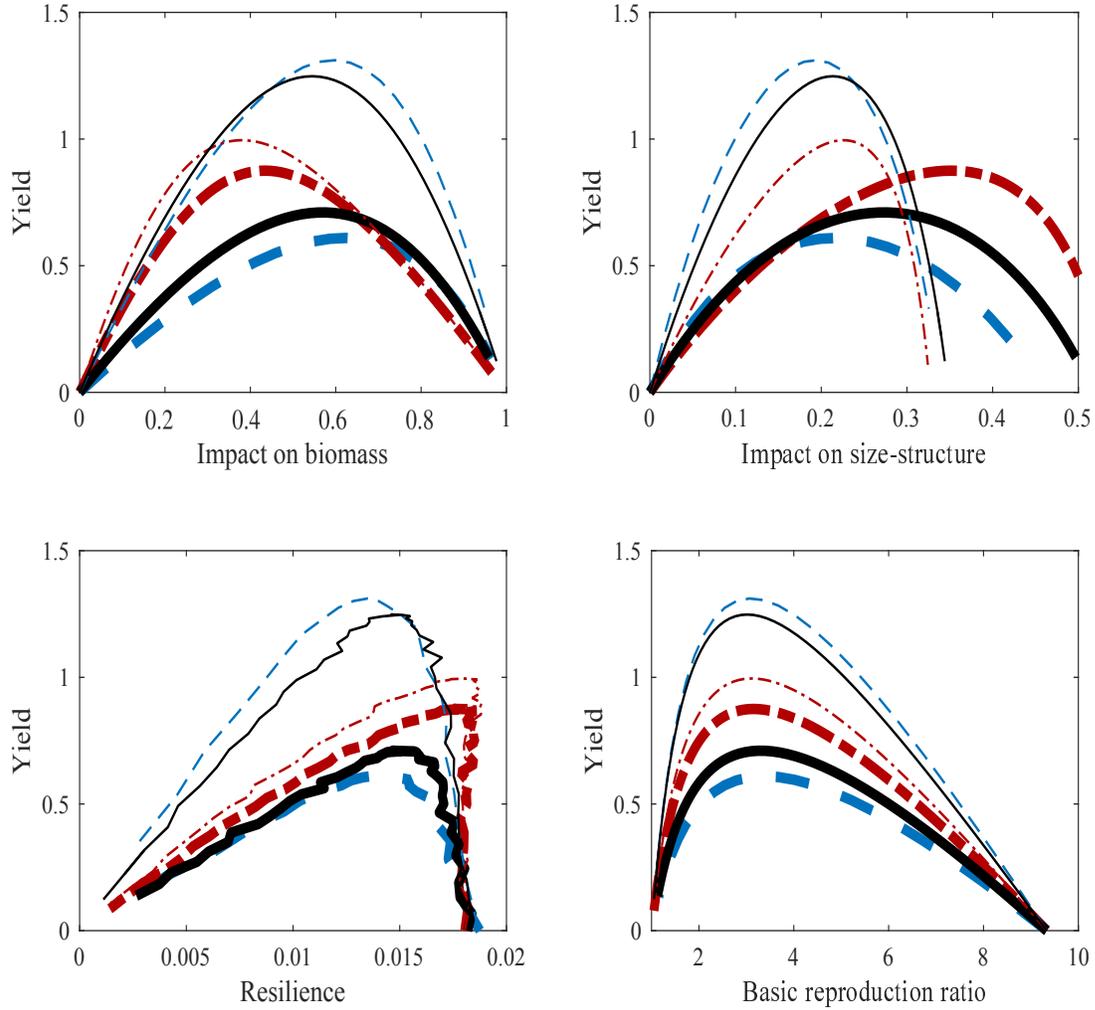}
\caption{
Age model for $a_0 = -3$ (thin curves) giving point $P_3$ in Fig. 4,
and $a_0 = -1$ (thick curves) giving point $P_4$.
Remaining description is as in Fig. A \ref{fig:age_steepness}.
}
\label{fig:age_a0}
\end{figure}

We present additional trade-off curves in Fig. A \ref{fig:age_a0} for
$a_0 = -3$ (point $P_3$ in Fig. 4) and $a_0 = -1$ (point $P_4$ in Fig. 4).
When $a_0 = -1$ then especially young individuals will have lower biomass and therefore the fraction of juveniles in the population will be small, $J^*_{\text{u}}/(J^*_{\text{u}} + A^*_{\text{u}}) \approx 53\%$.
As a result, adult harvesting performs better, with respect to all conservation measures,
than equal harvesting and juvenile harvesting.
When $a_0 = -3$ then especially young individuals will be of more biomass and therefore the fraction of juveniles in the population will be larger, $J^*_{\text{u}}/(J^*_{\text{u}} + A^*_{\text{u}}) \approx 66\%$.
In this case equal harvesting is on the Pareto front within the range of PGY for impact on biomass and resilience, and,
therefore we recommend equal harvesting.

Considering variations in the growth parameter $K$ gives a similar effect:
A large $K$ implies that individuals grow fast and this affects especially young individuals.
If $K = 1$ then $J^*_{\text{u}}/(J^*_{\text{u}} + A^*_{\text{u}}) \approx 81\%$ and juvenile harvesting is the only strategy on the Pareto front, in the range of PGY, for all conservation measures.
If $K = 0.1$ then $J^*_{\text{u}}/(J^*_{\text{u}} + A^*_{\text{u}}) \approx 41\%$ and adult harvesting is the only strategy on the Pareto front, in the range of PGY, for all conservation measures.

\subsection*{Age at maturity $a_{\text{mature}}$, point $P_5$ and $P_6$}

The quotient of length of first maturity to
maximum length of fish may vary in the wide range from 0.2 to 1,
but most species seem to be in the range 0.3 to 0.9
(Fishbase Fig. 43, http://www.fishbase.org/manual). 
Recalling the von Bertalanffy growth curves in Fig. A \ref{fig:growth-curve},
we realize that it is satisfactory that our results hold for age at maturity in the interval
$a_{\text{mature}} \in [3,15]$.

\begin{figure}
\centering
\includegraphics[height=14.5cm,width = 16cm]{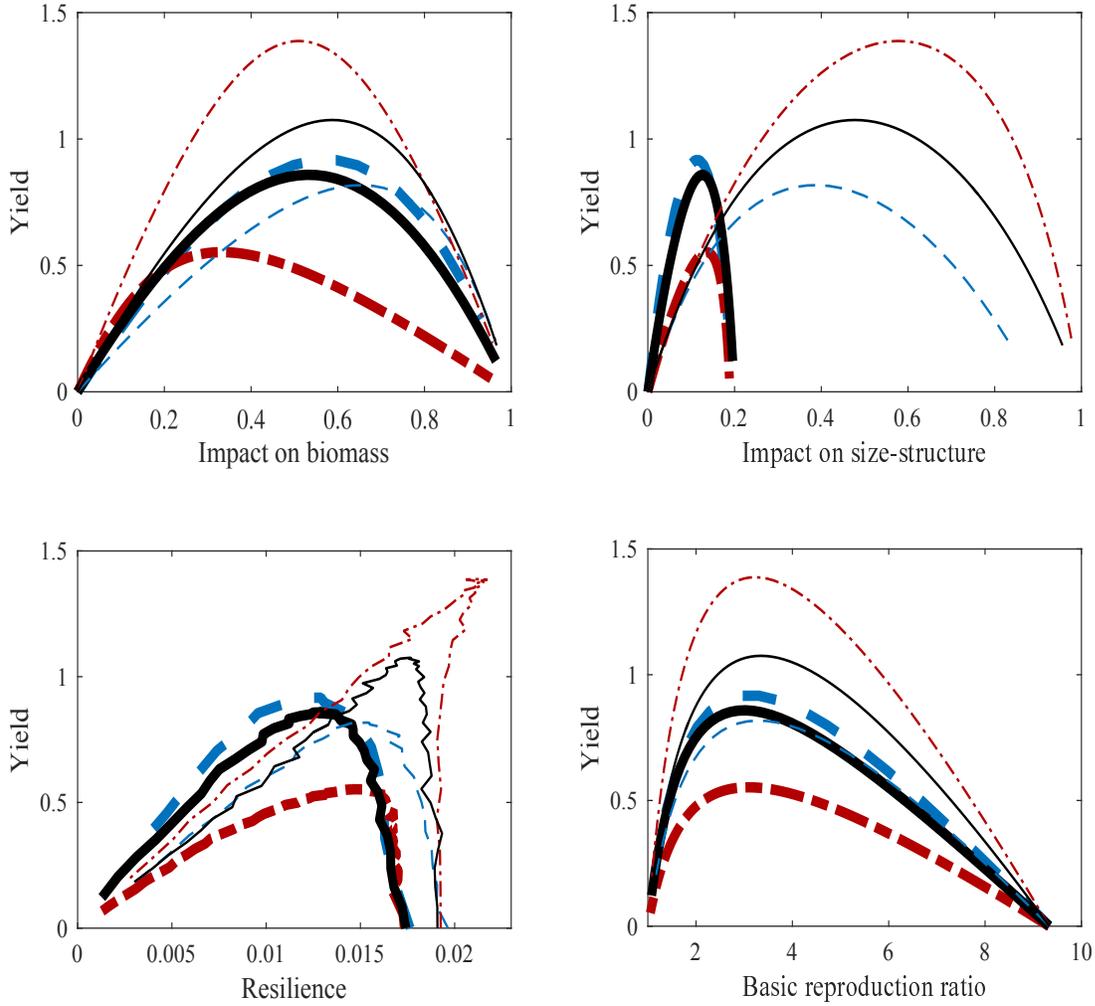}
\caption{
Age model for $a_{\text{mature}} = 5$ (thin curves) giving point $P_5$ in Fig. 4,
and $a_{\text{mature}} = 11$ (thick curves) giving point $P_6$.
Remaining description is as in Fig. A \ref{fig:age_steepness}.
}
\label{fig:age_amature}
\end{figure}

We present additional trade-off curves in Fig. A \ref{fig:age_amature}
for $a_{\text{mature}} = 5$ (point $P_5$ in Fig. 4) and $a_{\text{mature}} = 11$ (point $P_6$ in Fig. 4).
If $a_{\text{mature}} = 5$ then $J^*_{\text{u}}/(J^*_{\text{u}} + A^*_{\text{u}}) \approx 35\%$ and adult harvesting is the only strategy on the Pareto front, in the range of PGY, for all conservation measures.
If $a_{\text{mature}} = 11$ then $J^*_{\text{u}}/(J^*_{\text{u}} + A^*_{\text{u}}) \approx 78\%$ and we are at the border where equal harvesting and juvenile harvesting perform rather equally.
Thus, $a_{\text{mature}}$ has a large effect on the fraction of juveniles in the population, and,
therefore, a large impact on the preferable harvesting strategies.

\subsection*{Survival from natural mortality $S$, points $P_7$ and $P_8$}

\begin{figure}
\centering
\includegraphics[height=14.5cm,width = 16cm]{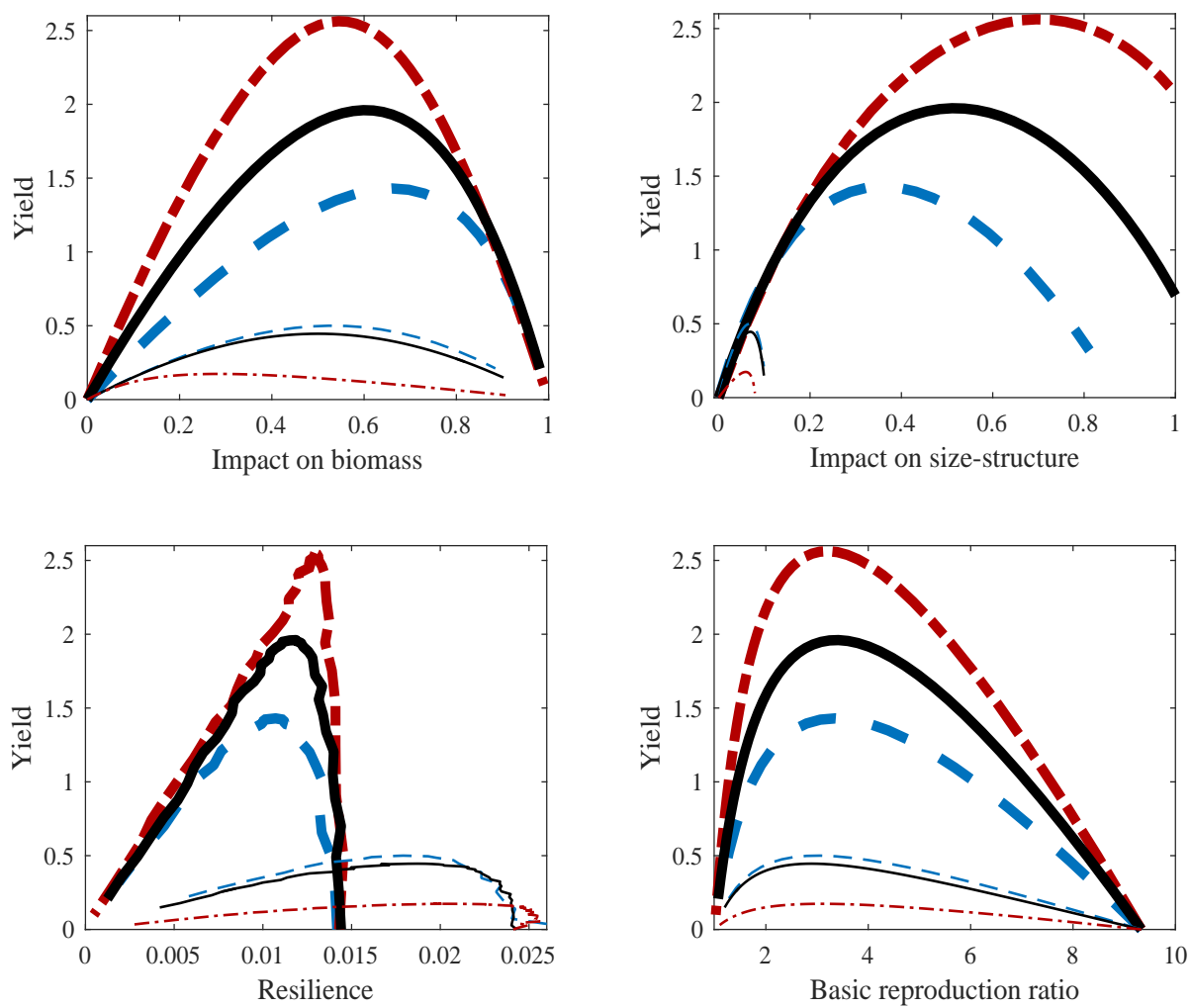}
\caption{
Age model for $S = 0.65$ (thin curves) giving point $P_7$ in Fig. 4,
and $S = 0.9$ (thick curves) giving point $P_8$.
Remaining description is as in Fig. A \ref{fig:age_steepness}.
}
\label{fig:age_S}
\end{figure}

Motivated by e.g. Mills et al. (2002) we used the default value $S = 0.8$ for the survival from natural mortality.
We have seen that results from the age model are rather dependent of $S$,
and this dependence has been investigated and presented in Fig. 4 in the main text.
Fig. A \ref{fig:age_S} shows additional trade-off curves for $S = 0.65$ (point $P_7$ in Fig. 4)
and $S = 0.9$ (point $P_8$ in Fig. 4).
If $S = 0.65$ then $J^*_{\text{u}}/(J^*_{\text{u}} + A^*_{\text{u}}) \approx 88\%$ and we are at the border where equal harvesting and juvenile harvesting perform rather equally.
If $S = 0.9$ then $J^*_{\text{u}}/(J^*_{\text{u}} + A^*_{\text{u}}) \approx 32\%$ and adult harvesting is the only strategy on the Pareto front, in the range of PGY, for all conservation measures.


\section*{Motivations and variations of parameter values in the stage model}
\label{sec:motiv_stage}

We adopt the following values of the model parameters,
\begin{align}\label{eq:Aparam_stage}\tag{A6}
H = T = r = 1, \quad R_{\text{max}}=2, \quad \sigma = 0.5, \quad I_{\text{max}} = 10, \quad M = 0.1, \quad z = 0.01, \quad q = 0.85.
\end{align}
These values may be considered as archetypal, and motivations can be found in de Roos et al. (2008).
However, these values were derived in the Roos et al. (2008) using not only data on fish species.
To ensure the validity of results in the setting of fisheries,
we give below a rather lengthy section concerning motivations and sensitivity of our results with respect to parameter values.

First, we recall that we have considered substantial variations of these values and concluded,
by numerically calculating trade-off curves, that our results from the stage model
are robust for the wide ranges of parameter values given by
\begin{align}\label{eq:Aparam_stage_interval}\tag{A7}
M \in [0, 0.5], \quad z \in [0.0001, 0.2], \quad  \sigma I_{\text{max}} \in [3,100], \quad q \in [0.6, 2], \quad R_{\text{max}} \in [0.5,100].
\end{align}
Indeed, we varied each parameter at a time,
keeping the others at the values given in \eqref{eq:Aparam_stage}
and tested at least 10 values in each interval.
Several further parameter combinations have also been tested, e.g. letting each parameter take on values at the boundary of the intervals in \eqref{eq:Aparam_stage_interval}.
The upper bound of 100 on the intervals for $R_{\text{max}}$ and $\sigma I_{\text{max}}$ does not seem necessary;
the structures of the population, relevant for our study, seem to remain for much larger values as well.
The lower limits of these intervals ensure that the population has a positive stable equilibrium.

In the following we will give,
in addition to the motivations in de Roos et al. (2008),
discussions and motivations for the parameter values in \eqref{eq:Aparam_stage}
as well as for the considered intervals in \eqref{eq:Aparam_stage_interval}.
We will also give some explanations of how and why our results depend, or not depend, on the parameter values.
Finally, we present additional trade-off curves
(similar to those in Fig. 3 in the main text) but for other parametrizations.
These curves are given in Figs. A \ref{fig:stage_z}, A \ref{fig:stage_q} and A \ref{fig:stage_d}.
Careful reading of these figures should convince the reader that our main findings
are robust with respect to variations of parameter values.
We do not present additional versions of main text Fig. 6 (showing relation between measures).
However,
the correlation can be seen from the trade-off figures by first focusing on the MSY point and then follow curves, for different measures,
in the direction of either increasing or decreasing harvesting pressure.

\subsection*{Reduction of parameters}

We begin by motivating that we can take $H = T = 1$ without loss of generality.
The half-saturation constant $H$ represents a resource density and is
measured in biomass per unit volume. 
Changing its value can be considered as changes in the volume in
which we express the densities $R$, $J$ and $A$.  
Without loss of generality we can choose $H = 1$.  
This fixes the environmental volume
in which the population is assumed to live and thus scales the biomass densities $R$, $J$ and $A$. 
Mathematically, we rescale the stage model by introducing 
the new dimensionless variables $J' = J/H$, $A' = A/H$, $R' = R/H$ and $R'_{\text{max}} = R'_{\text{max}}/H$,  
resulting in that $H$ goes away from the set of equations. 
We also rescale the time variable $t$ with the mass-specific metabolic rate parameter $T$, 
measured per unit of time, by introducing the dimensionless time $t' = t T$ and expressing the dynamics of the resource and
the consumers as functions of this new time variable. 
Moreover, we introduce the new dimensionless rate parameters $r' = r/T$, $I'_{\text{max}} = I_{\text{max}}/T$, $M' = M/T$, $F'_{\text{J}} = F_{\text{J}}/T$ and $F'_{\text{A}} = F_{\text{A}}/T$,
resulting in that $T$ goes away from the set of equations.
This shows that we can choose $H = T = 1$ without loss of generality,  
and that results for other values of the parameters $H$ and $T$ can be obtained by multiplications
from the results for $H = T = 1$. 
We therefore omit further investigations in $H$ and $T$ and put focus on the remaining parameters.
For simplicity, we will omit the ``prime"-notation in the following even though we work with the dimensionless quantities. 
The rescaled stage model equations are identical to main text eqs. (8) but with $H = T = 1$. 
In particular,   
\begin{align*}\label{eq:Astage-model}\tag{A8}
\frac{dJ}{dt} &= \left( w_{\text{J}}(R) - v(w_{\text{J}}(R)) - M - F_{\text{J}}\right) J  + 
w_{\text{A}}(R)A,\nonumber \\
\frac{dA}{dt} &= v(w_{\text{J}}(R))J - \left( M + F_{\text{A}}\right) A,\\
\frac{dR}{dt} &= r(R_{\max}-R)
-I_{\max}\frac{R}{1+R}\left(J+qA\right),\nonumber
\end{align*}
where the net biomass production by a juvenile and an adult are, respectively,
\begin{equation}\label{eq:Aproduction}\tag{A9}
w_{\text{J}}(R)= \max\left\{ 0, \sigma I_{\max} \frac{R}{1+R} - 1 \right\} \quad \text{and}\quad
w_{\text{A}}(R)= \max\left\{ 0, \sigma q I_{\max} \frac{R}{1+R} - 1 \right\},
\end{equation}
and the maturation is given by
\begin{align}\label{eq:Amaturation}\tag{A10}
v(x) = \frac{x - M - F_{\text{J}}} {1 - z^{1-(M + F_{\text{J}})/x}},
\end{align}
for $x \neq M + F_{\text{J}}$
and $v(M + F_{\text{J}}) = -(M + F_{\text{J}})/\log(z)$.

\begin{figure}
\centering
\includegraphics[height=6.5cm,width = 7.5cm]{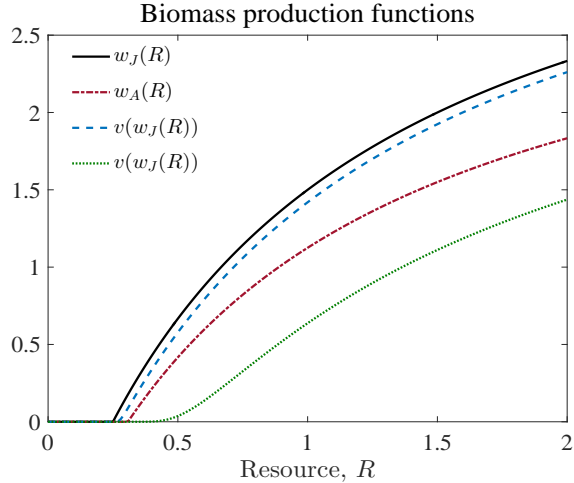}
\caption{Biomass production function for juveniles, $w_{\text{J}}(R)$, and for adults, $w_{\text{A}}(R)$.
The maturation function $v(w_{\text{J}}(R))$ for $M + F_{\text{J}} = 0.1$ (blue, dashed)
and for $M + F_{\text{J}} = 1$ (green, dotted).
Remaining parameters are as in \eqref{eq:Aparam_stage}.}
\label{fig:pedagogen}
\end{figure}

For clarity we plot the net biomass production functions for juveniles and adults,
$w_{\text{J}}(R)$ and $w_{\text{A}}(R)$,
as well as the maturation function $v(w_{\text{J}}(R))$, in Fig. A \ref{fig:pedagogen}.
All functions are zero (juveniles and adults do not produce biomass, and juveniles do not mature)
until energy intake is sufficient to cover maintenance requirements,
after which they increase with resource abundance.

\subsection*{Resource parameters $r$ and $R_{\text{max}}$}

Numerical investigations strongly indicate that our results are very robust with respect to variations of the resource parameters $r$ and $R_{\text{max}}$.
We will now give a theoretical explanation of this.
Let $J^*$, $A^*$ and $R^*$ denote the
juvenile, adult and resource biomass at equilibrium, respectively.
Let also $J_{\text{u}}^*$, $A_{\text{u}}^*$ and $R_{\text{u}}^*$
denote these quantities in case of no harvesting pressure,
i.e. when $F_{\text{J}} = F_{\text{A}} = 0$.
Setting derivatives to zero in the stage model \eqref{eq:Astage-model} and solving the corresponding set of equations for equilibria we find that
\begin{align}\label{eq:Abiomass_J_A}\tag{A11}
J^* = \frac{r(R_{\text{max}} - R^*)(1 + R^*)(M + F_{\text{A}})}
{I_{\text{max}} R^* (M + F_{\text{A}} + q v(w_{\text{J}}(R^*)))},\quad
A^* = \frac{r(R_{\text{max}} - R^*)(1 + R^*) v(w_{\text{J}}(R^*))}
{I_{\text{max}} R^* (M + F_{\text{A}} + q v(w_{\text{J}}(R^*)))},
\end{align}
where $R^*$ is given as the unique root of the equation
\begin{align}\label{eq:Abiomass_R}\tag{A12}
\frac{w_{\text{A}}(R^*) v(w_{\text{J}}(R^*))}{(M + F_{\text{A}}) [v(w_{\text{J}}(R^*)) - w_{\text{J}}(R^*) + M + F_{\text{J}}]} = 1.
\end{align}
%
%
From \eqref{eq:Abiomass_R} we note that the root $R^*$ will be independent of both $r$ and $R_{\text{max}}$,
and from \eqref{eq:Abiomass_J_A} we see that both the juvenile and adult biomass, $J^*$ and $A^*$,
increase linearly with both $r$ and $R_{\text{max}}$, and so is the yield.
Moreover, from the expressions in \eqref{eq:Abiomass_J_A} 
we see that
\begin{align}\label{eq:Ajohejsats1}\tag{A13}
\frac{J^*}{J^* + A^*} &\,=\, \frac{M + F_{\text{A}}}{M + F_{\text{A}} + v(w_{\text{J}}(R^*))}
\quad \text{and}\notag\\
\text{Impact on size-structure}
&\,=\, \frac{M + F_{\text{A}}}{M + F_{\text{A}} + v(w_{\text{J}}(R^*))}
\left[ \frac{M}{M + v(w_{\text{J}}(R_{\text{u}}^*))}\right]^{-1} - 1.\nonumber
\end{align}
Therefore, the fraction of juveniles in the population, as well as the population size structure,
are independent of $r$ and $R_{\text{max}}$ (because $R^*$ is).
We can also conclude,
from the expressions for the recovery potential and the impact on biomass given in the main text,
that both these measures are independent of $r$.
We may also note that the recovery potential becomes independent of $R_{\text{max}}$ as $R_{\text{max}}$ becomes very large as the ingestion for a single individual saturates for very large resource densities.

These arguments clarifies why our results are more or less independent of $r$ and $R_{\text{max}}$
(whenever these parameters take on values giving a non-extinct population at equilibrium).
Therefore, we feel comfortable with letting $r = 1$ and $R_{\text{max}} = 2$ and put focus on the remaining parameters.

\subsection*{Ingestion parameters $\sigma$ and $I_{\text{max}}$}

Numerical investigations strongly indicate that our results are very robust with respect to variations of
the ingestion parameters $\sigma$ and $I_{\text{max}}$.
To understand why this is the case,
we expand on the above analytical reasoning and prove that at equilibrium,
the fraction of juvenile biomass in the population and the impact on size structure,
see \eqref{eq:Ajohejsats1}, as well as $w_{\text{A}}(R^*)$, $w_{\text{J}}(R^*)$ and $v(w_{\text{J}}(R^*))$
are in fact independent of $r$, $R_{\text{max}}$, $\sigma$ and $I_{\text{max}}$.
Indeed, they depend only on the parameters
$M + F_{\text{J}}$, $M + F_{\text{A}}$, $q$, and $z$.
To prove this, consider equation \eqref{eq:Abiomass_R} from which we solve the resource equilibrium biomass.
In this equation, $R^*$, $\sigma$ and $I_{\max}$ appear only implicitly in the functions
$w_{\text{A}}(R^*)$ and $w_{\text{J}}(R^*)$, see expressions \eqref{eq:Aproduction}.
Thus, by setting
\begin{align*}
\alpha^* = \sigma I_{\max} \frac{R^*}{1+R^*}
\end{align*}
we can solve \eqref{eq:Abiomass_R} for $\alpha^*$,
independent of the parameters $\sigma$ and $I_{\max}$.
Therefore, varying $\sigma$ and $I_{\max}$ will not change
$w_{\text{A}}(R^*)$, $w_{\text{J}}(R^*)$ and $v(w_{\text{J}}(R^*))$,
and thus not change the fraction of juvenile biomass in the population nor the impact on size structure.
This shows why our results are very robust with respect to $\sigma$ and $I_{\max}$,
and from here we fix $\sigma = 0.5$ and $I_{\max} = 10$ and focus on the remaining parameters.

\subsection*{Ratio of size at birth to maximum size $z$}

De Roos et al. (2008) use the value $z = 0.01$ for the ratio of size at birth to maximum size,
$z = s_{\text{born}} / s_{\text{max}}$.
To further motivate this value in a fisheries context,
we note that Punt et al. (1995, page 290)
studying the albacore (\textit{Thunnus alalunga}, Scombridae) use $K = 0.23$ and $a_0 = -1$
in the von Bertalanffy growth growth curves, recall Fig. A \ref{fig:growth-curve}.
This gives a value of $z \approx 0.009$ which is very close to 0.01.
(Changing $a_0$ to $-3$ gives $z \approx 0.12$, and $a_0 = -0.2$ gives $z \approx 0.00009$.)

\begin{figure}
\centering
\includegraphics[height=14.5cm,width = 16cm]{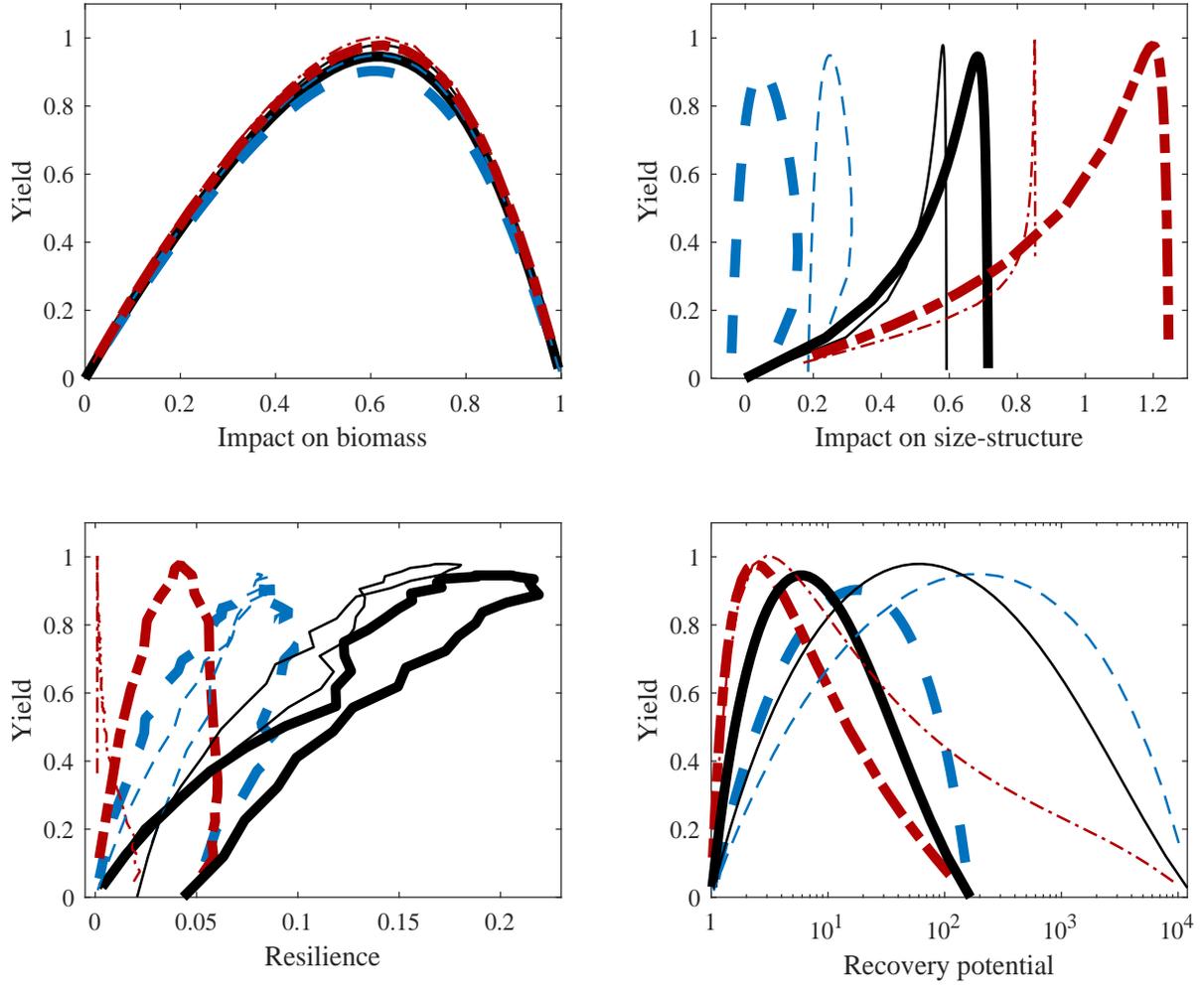}
\caption{Stage model for $z = 0.001$ (thin curves) and $z = 0.1$ (thick curves).
Juvenile harvesting (blue, dashed), equal harvesting (black, solid), adult harvesting (red, dash-dot).
Remaining parameters are as in \eqref{eq:Aparam_stage} and
yield is normalized as in Fig. 3 in the main text.
}
\label{fig:stage_z}
\end{figure}

By numerically calculating trade-off curves as those in Fig. 3 in the main text,
we have seen that our results are robust with respect to variations of $z$ in the wide interval $z \in [0.0001, 0.2]$.
Indeed, results from the stage model suggest that equal harvesting performs well rather independent
of $z$ as long as it takes on reasonable values.
We present additional trade-off curves in Fig. A \ref{fig:stage_z} for the values $z = 0.001$ and $z = 0.1$.
Here, we clearly see that the value of $z$ has little effect on how juvenile-, adult- and equal harvesting performs.

\subsection*{Difference in ingestion between juveniles and adults $q$}

\begin{figure}
\centering
\includegraphics[height=14.5cm,width = 16cm]{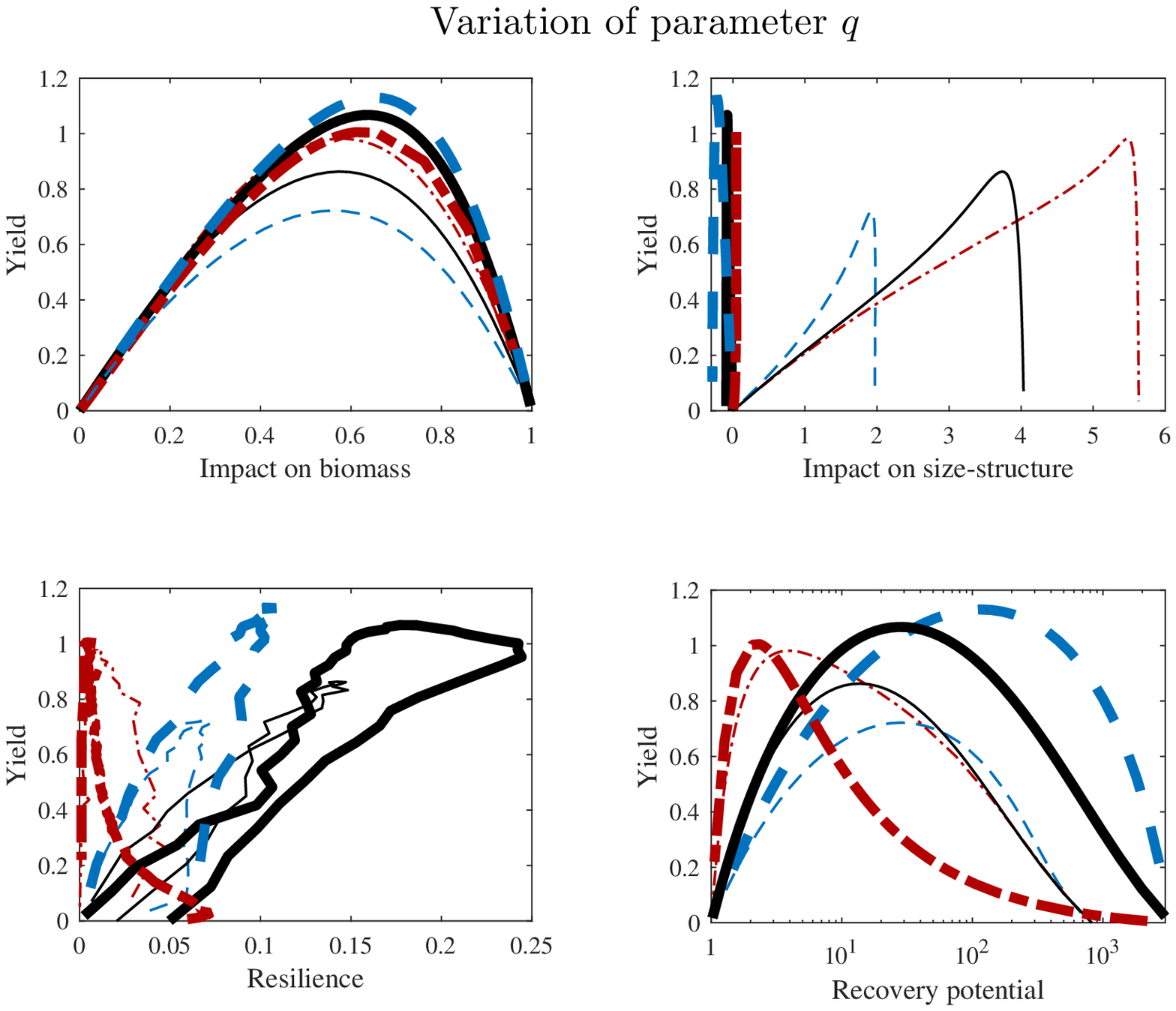}
\caption{
Stage model for $q = 0.6$ (thin curves) and $q = 1.5$ (thick curves),
remaining description is as in Fig. A \ref{fig:stage_z}.
}
\label{fig:stage_q}
\end{figure}

It seems hard to motivate a specific value for the parameter $q$ describing the difference in
ingestion rates between juveniles and adults:
``An estimate for this parameter can hardly be derived from
experimental data as it is only a phenomenological
representation of stage-specific differences in resource
availability and resource use between juveniles and adults." (De Roos et al. 2008).
By substantial numerical investigations we have seen that our results hold at least as long as $q \in [0.6, 2]$.
We present additional trade-off curves in Fig. A \ref{fig:stage_q} for $q = 0.6$ and $q = 1.5$,
showing that our main findings hold for these parametrizations as well.
The conservation measures stays rather stable also when $q$ is goes out of the bounds 0.6 and 2,
but then the difference in yield between the harvesting strategies increase.
Indeed, smaller $q$ implies that adult harvesting is better,
and larger $q$ implies that juvenile harvesting is better.


%

\subsection*{Natural mortality rate $M$}

\begin{figure}
\centering
\includegraphics[height=14.5cm,width = 16cm]{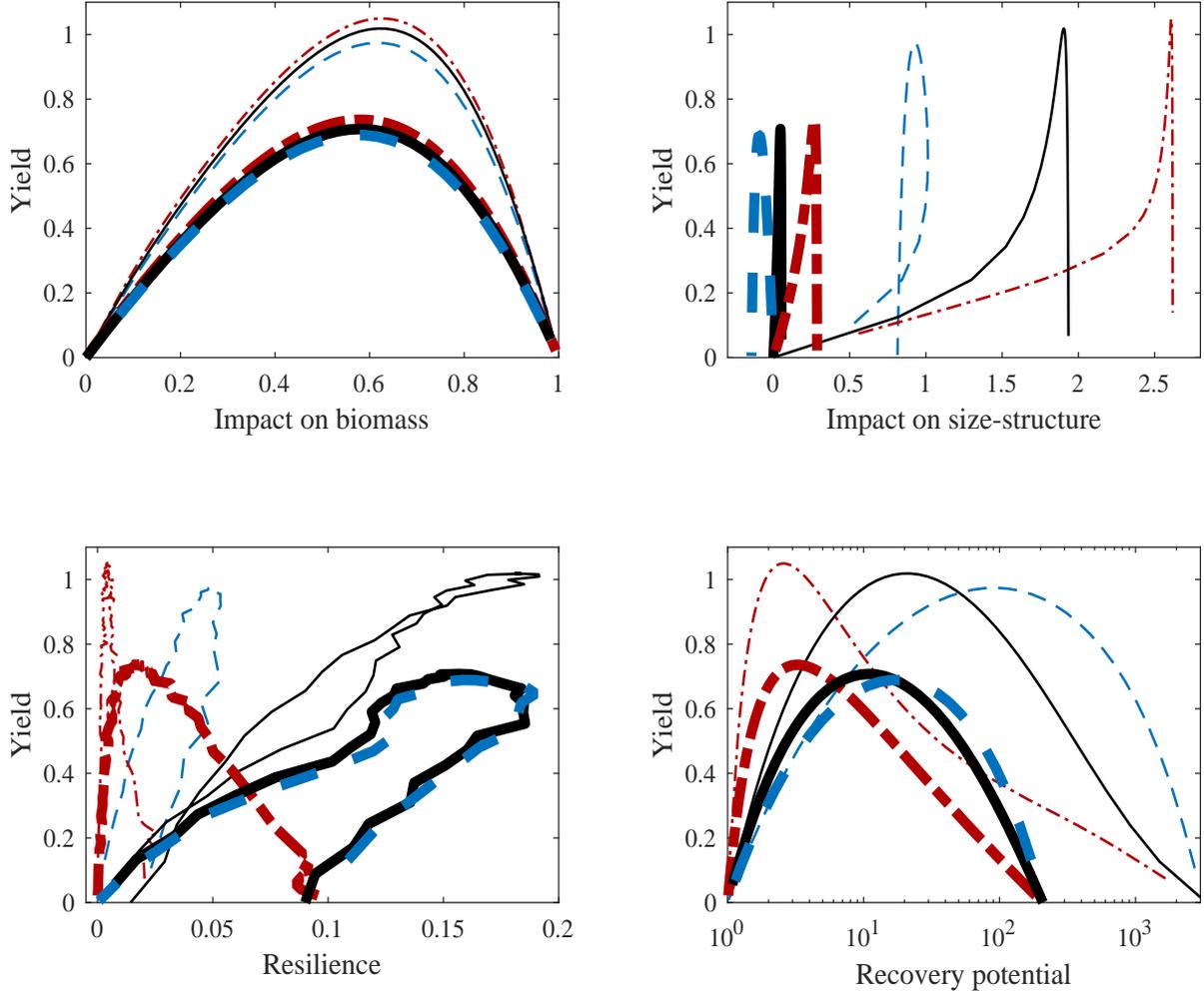}
\caption{
Stage model for $M = 0.05$ (thin curves) and $M = 0.4$ (thick curves),
remaining description is as in Fig. A \ref{fig:stage_z}.
}
\label{fig:stage_d}
\end{figure}

Following De Roos et al. (2008) we have adopted the default value of $M = 0.1$ for the natural mortality rate.
Fig. A \ref{fig:stage_d} shows additional trade-off curves for the values $M = 0.05$ and $M = 0.4$.
The main structure of the curves remains and thus our main results are robust also to changes in $M$.
Let us note that $M$ is simply added to the harvesting rates $F_{\text{J}}$ and $F_{\text{A}}$ in the stage model,
recall \eqref{eq:Astage-model}.
Therefore, an increase in $M$ should force the trade-off curves for juvenile harvesting and adult harvesting toward the trade-off curves for equal harvesting
(if $M$ is high, then there is then always high mortality on both juveniles and adults).
This can be observed in Fig. A \ref{fig:stage_d}.
We can also see that the impact on size structure is much smaller for higher $M$,
which may be understood by the fact that
now the unharvested state already has a high death rate.
We can also observe the expected decrease in yield when death rate increases from $M = 0.05$ to $M = 0.4$.




\section*{Detailed description of our resilience measure}

We have considered \textit{resilience} of the population by measuring the reciprocal of the time needed
for the population to recover the positive equilibrium given a random perturbation.
To describe the procedure in detail,
let $(J_{\text{}}^*, A_{\text{}}^*, R_{\text{}}^*)$ denote the equilibrium of biomass in the stage model,
and let $N_{a}^*$, $a = 0,1,2, \dots, a_{\text{max}}$ denote the equilibrium of number of individuals in the age model.
For a given constant $\kappa > 0$ that scales the maximum displacement of the population from the equilibrium, we start a trajectory from a random point uniformly distributed in the cube
\begin{align}\label{eq:Aperturb-imposed}\tag{A14}
(0, \kappa J_{\text{}}^* ) \times (0, \kappa A_{\text{}}^*) \times (0, \kappa R_{\text{}}^*) \qquad \text{(stage model)},&\\
(0, \kappa N_{0}^* ) \times (0, \kappa N_{1}^*) \times (0, \kappa N_{2}^*) \times \cdots \times (0, \kappa N_{a_{\text{max}}}^*) \qquad \text{(age model)}&. \notag
\end{align}
We then find the return time as the time needed for this trajectory to be close enough to
the equilibrium in the sense that
%
%
%
%
%
\begin{align}\label{eq:Aneighbor}\tag{A15}
\left\{ \left(\frac{J_{\text{}}^* - J(t)}{J_{\text{}}^*}\right)^2
+ \left(\frac{A_{\text{}}^* - A(t)}{A_{\text{}}^*}\right)^2\right.
+ \left.\left(\frac{R_{\text{}}^* - R(t)}{R_{\text{}}^*}\right)^2\right\}^{1/2} \leq \epsilon \qquad &\text{(stage model)},\\
\left\{\sum_{a = 0}^{a_{\text{max}}} \left(\frac{N_{a}^* - N_{a,t}}{N_{a}^*}\right)^2\right\}^{1/2} \leq \epsilon  \qquad &\text{(age model)},\notag
\end{align}
for some small $\epsilon > 0$.
After repeating this procedure $n$ times
we find the resilience,
as a function of the harvesting rates $F_{\text{J}}$ and $F_{\text{A}}$,
by taking the reciprocal of the average of the corresponding return times, i.e.
\begin{align*}
\text{Resilience} \,=\, \frac{1}{\text{Average of the $n$ return times}}.
\end{align*}
%
Our resilience measure estimates the population's expected rate of return, given a random perturbation.
We have used the parameter values $\kappa = 2$, $\epsilon = 0.01$ and $n = 100$ as default values,
but main results are not very sensitive to the magnitude of the imposed perturbations ($\kappa$) nor the size of the
small neighborhood ($\epsilon$) as long as $\kappa >> \epsilon$ take on reasonable values.
We present some results on $\kappa = 1$ in Fig. A \ref{fig:resil-kappa1} below,
which means that we consider only perturbations that remove biomass.

It is worth noting that when implementing our approach to resilience, 
one tests the model for a large number of initial conditions.
Therefore, one may find out if the system is bistable, i.e. if other attractors (e.g. stable states or periodic solutions) are present,
which is valuable information. 
Indeed, if this is the case, a tested initial condition may produce a trajectory that never return to the equilibrium. 
This will be revealed by our simulation procedure. 
If applying our approach to bistable systems, one should define another stopping criteria for the other attractors. 
The resilience can then be defined as the average of the reciprocal of the return times, setting the reciprocal to zero whenever an initial condition reached another attractor.
In that way initial conditions (perturbations) that escape to another states contributes with resilience zero, 
c.f. measure $\mathcal{R}$ in Lundstr\"om (2018).

\section*{Basin-time resilience and further investigations of the stage model}

From our results on the stage model we have seen that resilience,
in contrary to the other conservation measures,
may vastly increase with harvesting pressure for some harvesting strategies.
This is true for equal harvesting and juvenile harvesting,
but not at all for adult harvesting,
recall Fig. 3 in the main text as well as Figs. A \ref{fig:stage_z}, A \ref{fig:stage_q} and A \ref{fig:stage_d}.
Indeed, adult harvesting gives the worst resilience while equal harvesting gives the best resilience
(which can be up to 20 times higher in the range of PGY).
As this phenomenon provides a substantial argument in favour of equal harvesting compared to adult harvesting,
we investigate it further here in order to increase the credibility of our results.

\begin{figure}
\centering
\includegraphics[height=6.5cm,width = 7.5cm]{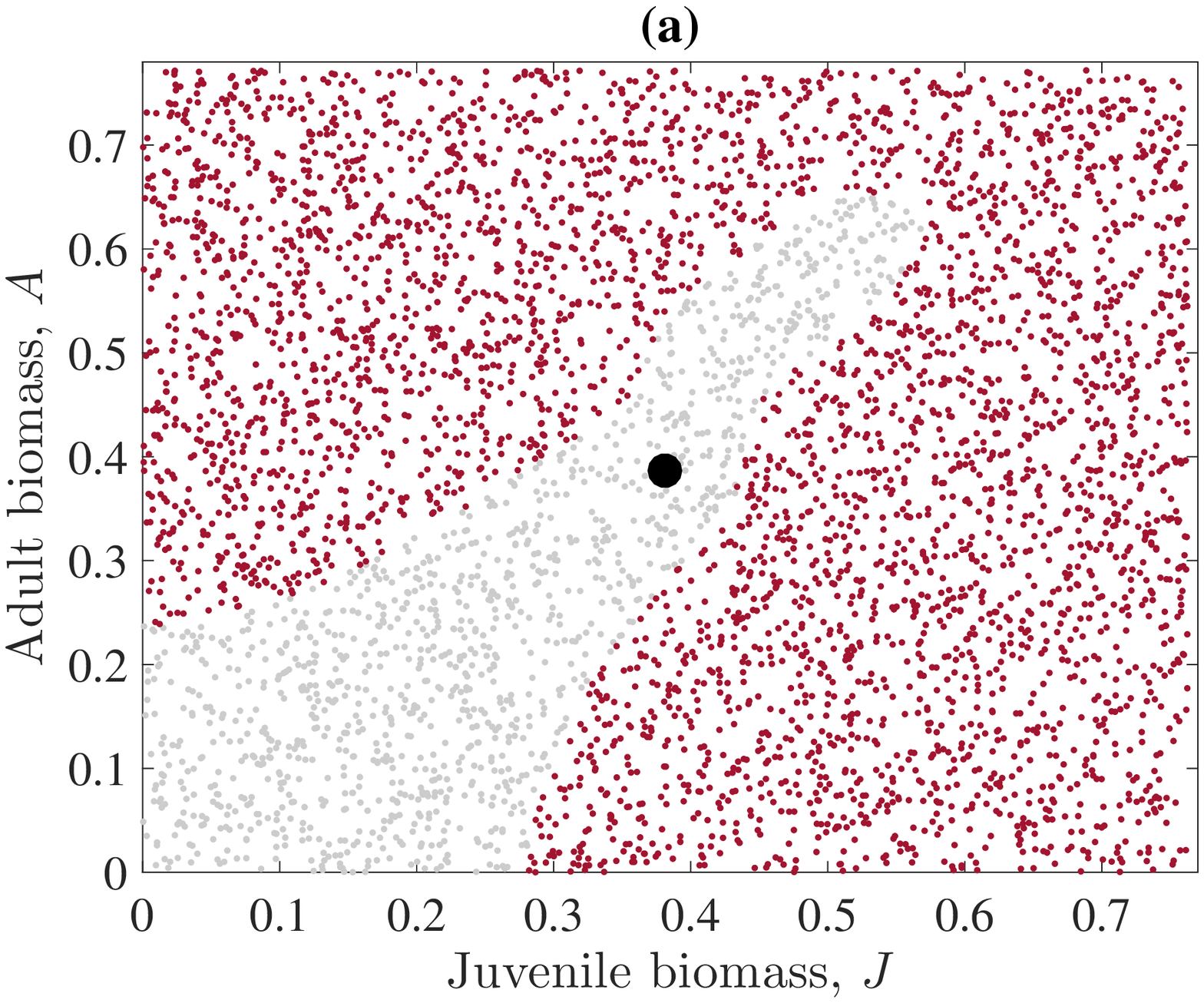}\hspace{0.1cm}
\includegraphics[height=6.5cm,width = 7.5cm]{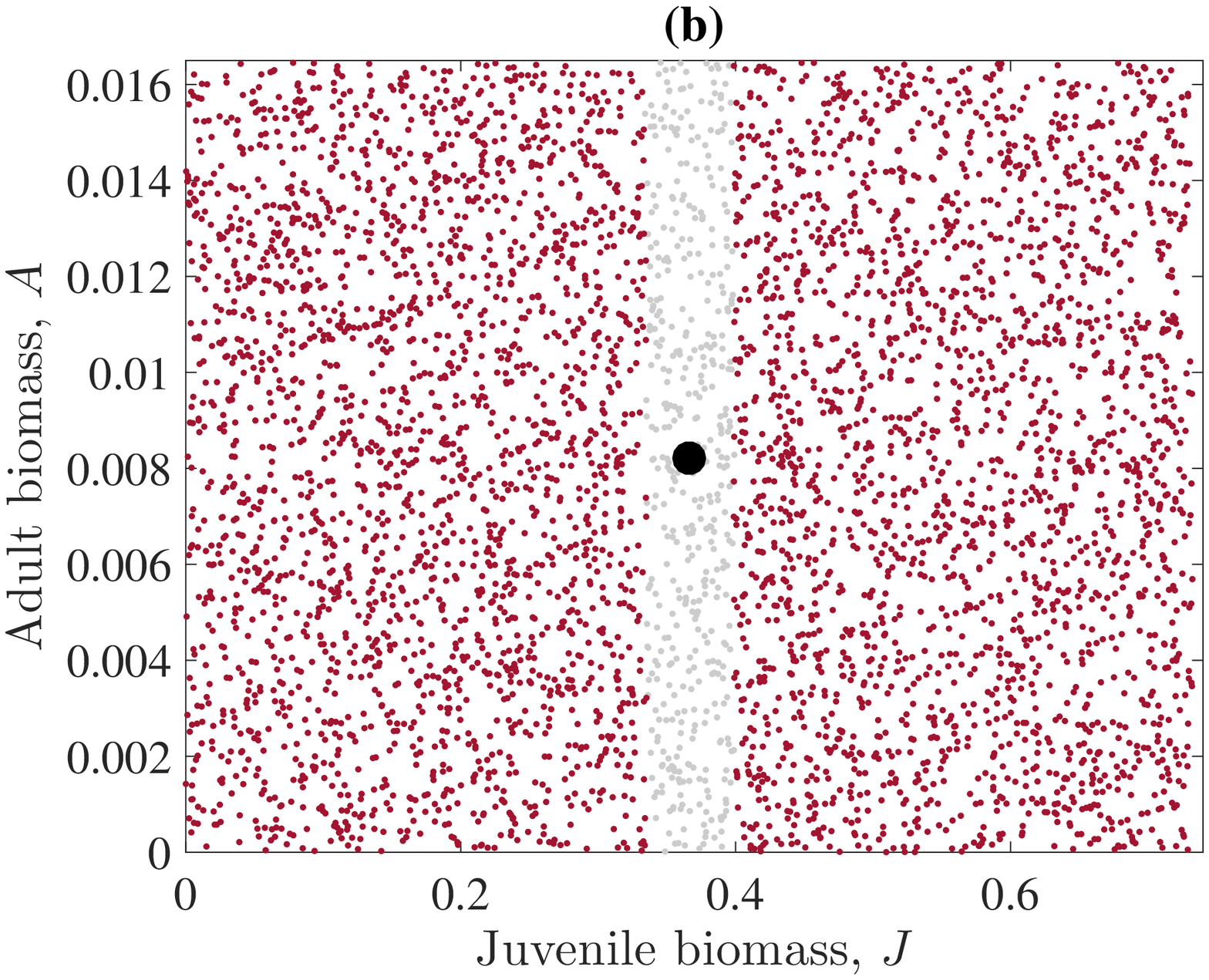}
\includegraphics[height=6.5cm,width = 7.5cm]{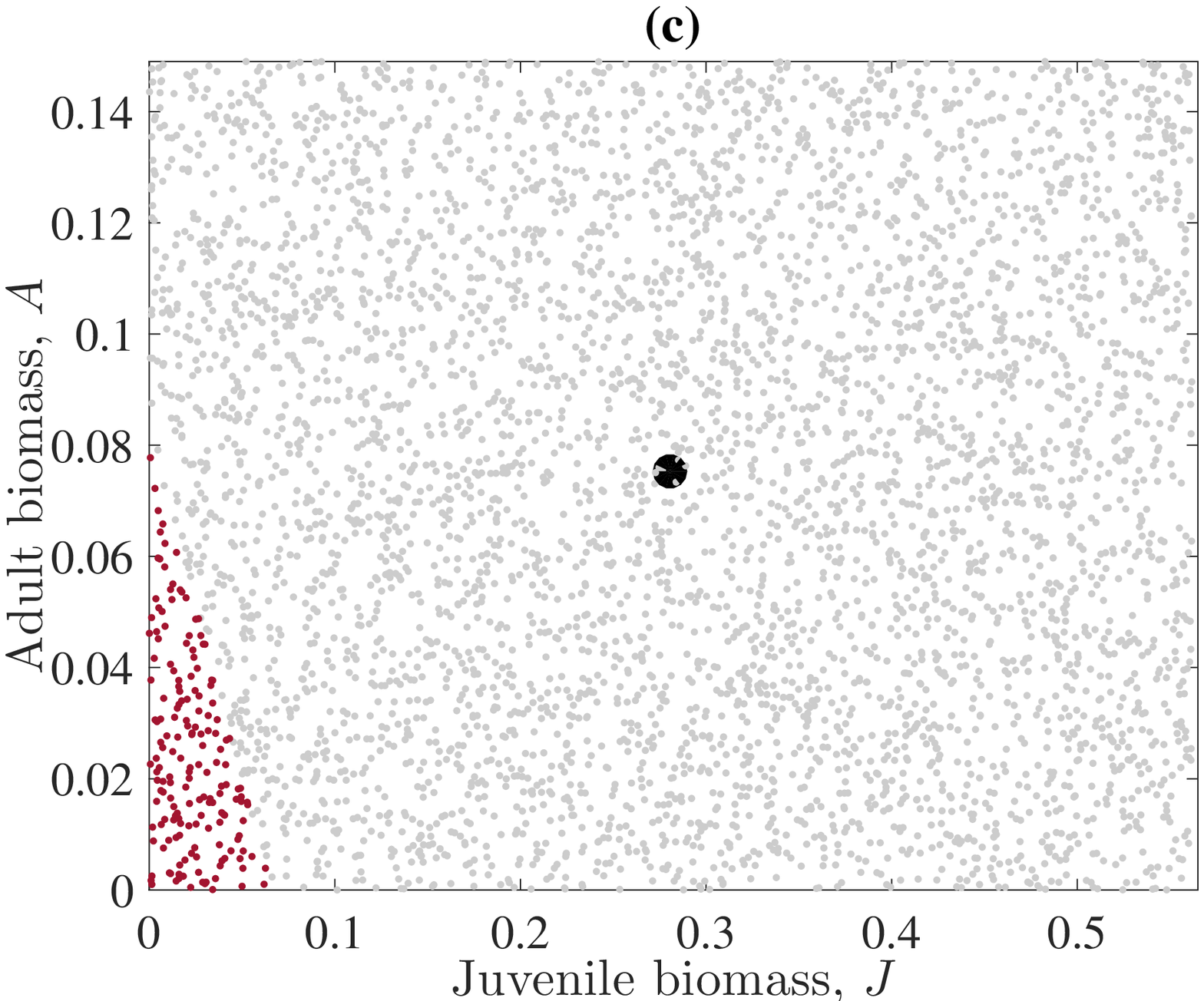}\hspace{0.3cm}
\includegraphics[height=6.5cm,width = 7.5cm]{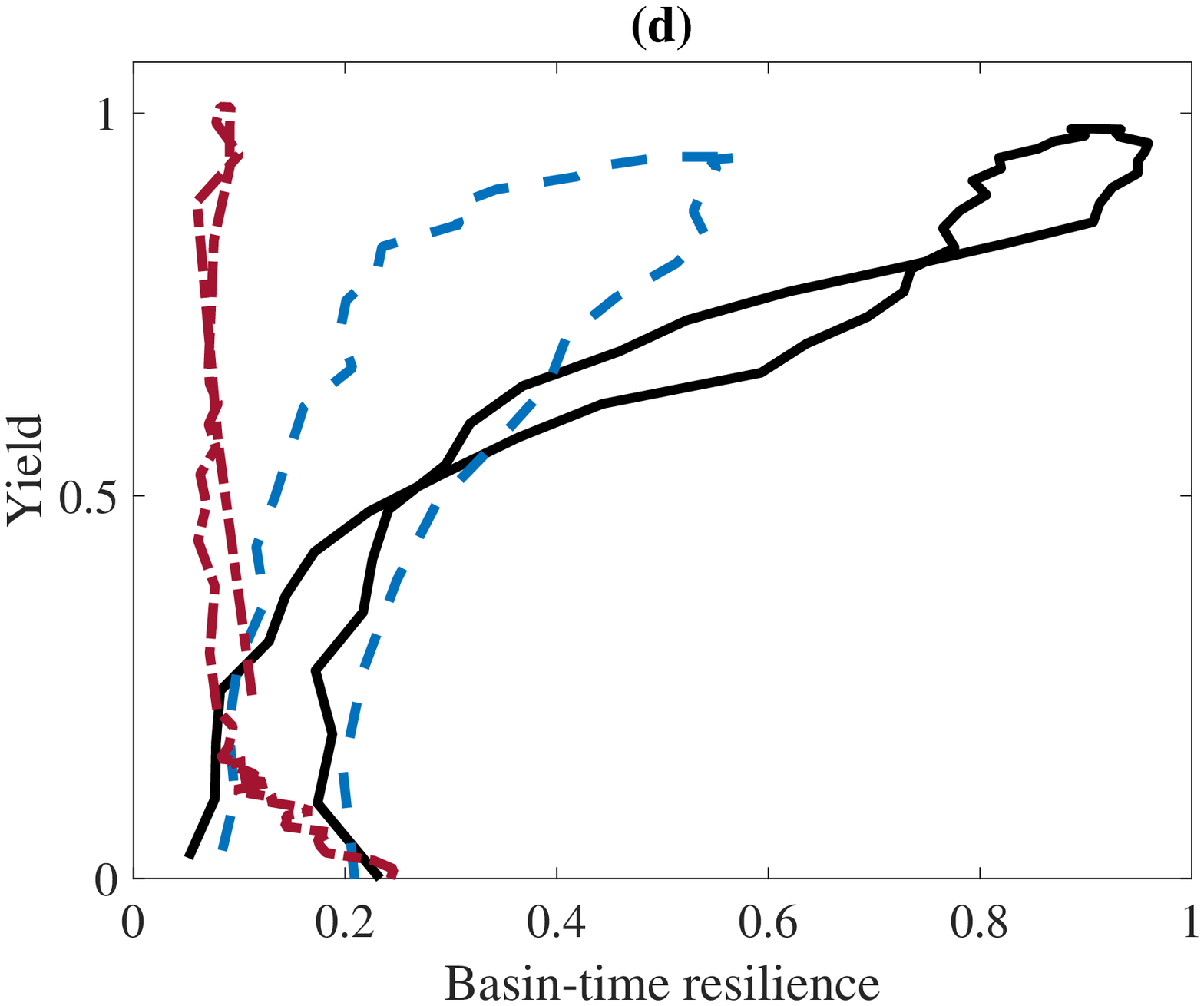}
\caption{(a)-(c): Population recover fast from most perturbations in case of equal harvesting (c) and from fewest perturbations in case of  adult harvesting (b).
The grey dots represent initial conditions that return to the equilibrium within time $5$,
while the red dots need more time to recover the equilibrium.
There is no harvesting pressure in (a).
(d): Trade-off between basin-time resilience and yield.
Juvenile harvesting (blue, dashed), equal harvesting (black, solid), adult harvesting (red, dash-dot).
Model parameters are as in \eqref{eq:Aparam_stage} and
yield is normalized as in Fig. 3 in the main text.
}
\label{fig:johej-ny}
\end{figure}

Each dot in Figs. A \ref{fig:johej-ny} (a)-(c) show an imposed perturbation as an initial condition from which we integrated the trajectory until it reached the small neighborhood of the equilibrium given by \eqref{eq:Aneighbor} with $\epsilon = 0.1$.
Tested initial conditions are distributed uniformly in
$(0, \kappa J^*) \times (0, \kappa A^*)$ with $R = R^*$ and $\kappa = 2$.
Grey dots return to the equilibrium within a time limit which we set to $t \leq 5$,
while red dots need more time to recover the equilibrium.
There is no harvesting pressure in Fig. A \ref{fig:johej-ny} (a),
adult harvesting $F_J = 0, F_A = 35$ in (b) and equal harvesting $F_J = F_A = 0.8$ in (c).
The harvesting strategies in (b) and (c) give similar yields (close to the maximum possible yield).
The fraction of grey dots is largest in case of equal harvesting,
while it is smallest in case of adult harvesting, which
means that the population returns fast from more perturbations (initial conditions)
in case of equal harvesting than for adult harvesting.
This reflects the higher resilience in case of equal harvesting.
Indeed, the fraction of grey dots estimates the probability that the population recover the equilibrium in a time limit,
given a random perturbation, and constitutes a natural candidate for measuring resilience.

In general, let \emph{basin-time}
be a subset of the basin of attraction from which trajectories
return to (a small neighborhood of) the attractor within a time limit, $t \leq \tau$.
A large and convex basin-time set should then reflect high resilience.
Therefore, both the size and the shape of the basin-time constitute natural candidates for measuring resilience \citep{jag0}.
We will consider the size of the basin-time set as a resilience measure.
To do so we impose perturbations as described in \eqref{eq:Aperturb-imposed} and
integrate each trajectory until it reaches the neighborhood defined by \eqref{eq:Aneighbor}.
We further let $N_{\text{return}}^{\tau}$ denote the number of initial conditions from which trajectories
return in time $\tau$, and $N_{\text{tot}}$ denote the total number of tested initial conditions.
We measure resilience as the size of the basin-time set through the estimate
\begin{align*}
\text{Basin-time resilience}\, =\, \frac{N_{\text{return}}^{\tau}}{N_{\text{tot}}}.
\end{align*}
Figure A \ref{fig:johej-ny} (d) shows trade-off curves between basin-time resilience and yield.
(parameters are $\kappa = 2$, $n = 1000$, $\epsilon = 0.1$ and $\tau = 5$.)
It is clear that also the basin-time resilience measure strengthen the result of the stage model saying 
that equal harvesting performs best while adult harvesting gives the worst resilience;
these trade-off curves are very similar to those in Fig. 3 (c) in the main text.

\begin{figure}
\centering
\includegraphics[height=6.5cm,width = 7.5cm]{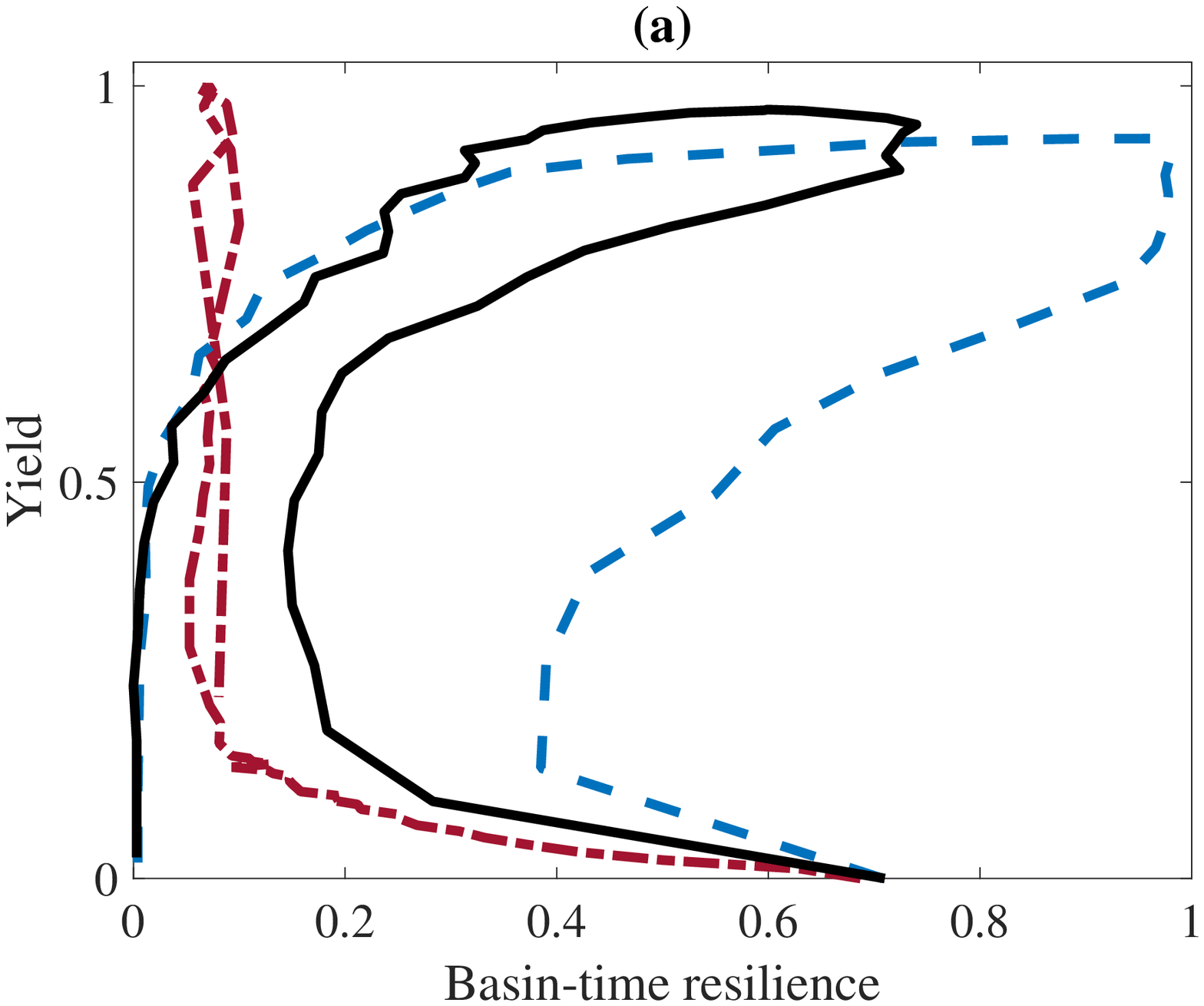}\hspace{0.1cm}
\includegraphics[height=6.5cm,width = 7.5cm]{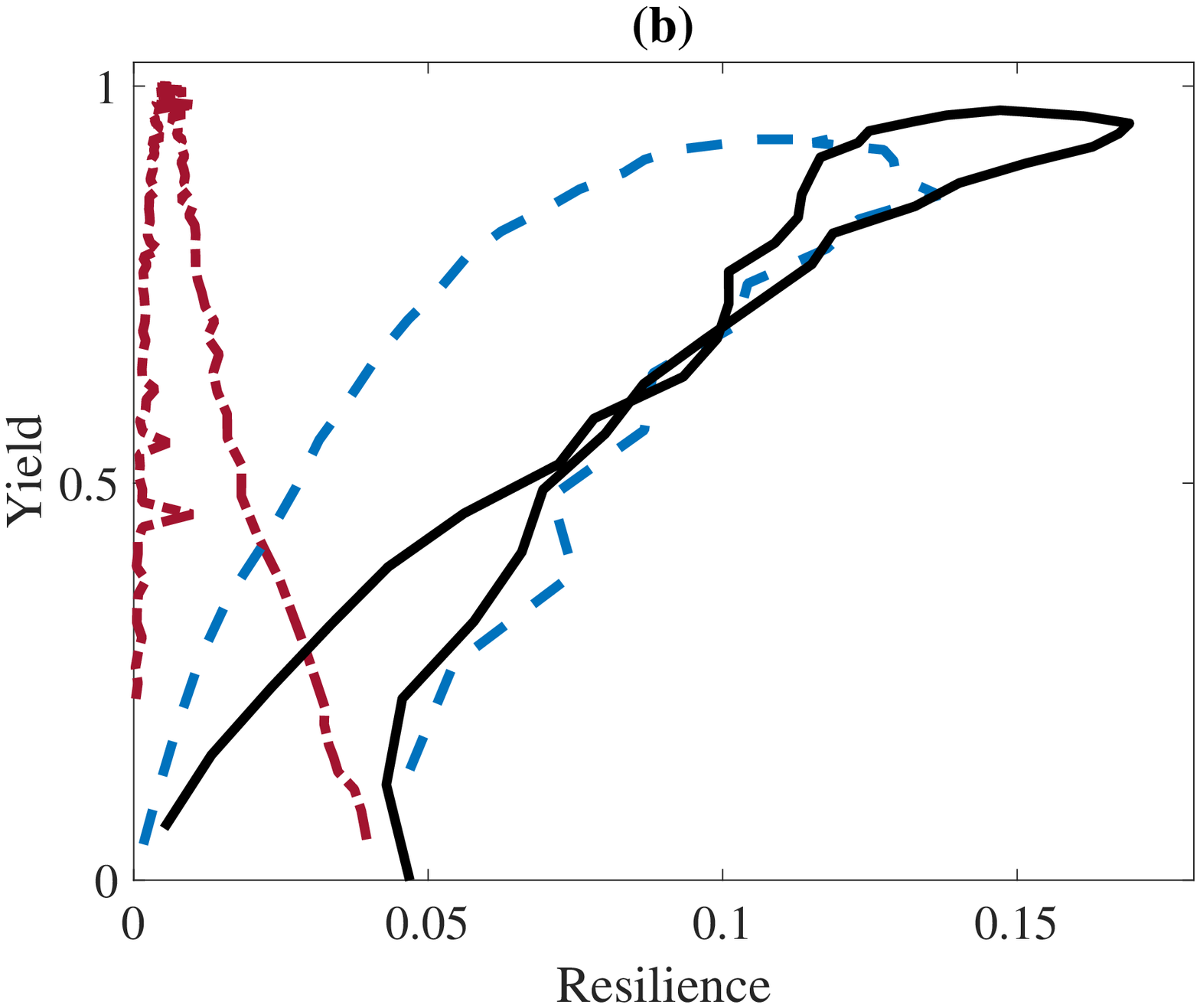}
\caption{Trade-off between yield and resilience considering $\kappa = 1$, i.e. perturbations only remove population and resource biomass. (a) basin-time resilience and (b) the resilience measure used in the main text.
Remaining descriptions are as in Fig. A \ref{fig:johej-ny}.
}
\label{fig:resil-kappa1}
\end{figure}

Figs. A \ref{fig:johej-ny} (a), (b) and (c) may indicate that the higher resilience in case of equal harvesting is due to fast return of trajectories starting at high biomass.
Indeed, the basin-time, estimated by the fraction of grey dots, is largest in Fig. A \ref{fig:johej-ny} (c) and includes all points corresponding to larger initial biomass than at equilibrium.
Therefore, we complement the given resilience estimations by a simulation in which we consider
only perturbations that remove biomass.
In particular, we take $\kappa = 1$ and reproduce the trade-off curves between resilience (both measures) and yield,
see Fig. A \ref{fig:resil-kappa1}.
We can observe that equal harvesting is now challenged by juvenile harvesting,
but both resilience measures still agree on the result that equal harvesting is much better than adult harvesting.


In addition to the above resilience estimations,
we have also considered normally distributed perturbations (centered at the equilibrium)
as well as the ``standard" local resilience measure
given by the real part of the eigenvalue of the Jacobian matrix at equilibrium with largest real part.
Results from these investigations also agree very well with our findings.

For further ideas on how to use the basin of attraction,
the basin-time as well as the return time to measure stability and resilience,
we refer the reader to Lundstr\"om (2018) and references therein.


\section*{Derivation of the basic reproduction ratio for the age model}
%
The basic reproduction ratio represents the average number of offsprings produced over the
lifetime of an individual in the absence of density-dependent competition.
We consider an individual satisfying the dynamics of the age model 
and calculate an expression for this individuals expected offsprings during its lifetime.
For this purpose,
we let $\tilde{E}_t$ denote the egg production and $\tilde{R}_t$ the recruitment from this individual,
at time $t$.
That no density-dependent competition is present lead us to assume that
the Beverton-Holt recruitment is completely unsaturated.
Such situation is achieved by taking $\beta = 0$ (or $\mathcal{R}_0 = \infty$) in \eqref{eq:Arecruitment},
giving
\begin{align}\label{eq:Aproof1}\tag{A16}
\tilde{R}_{t+1} = \frac{\tilde E_t}{\alpha} \,\text{exp} \left( u_t- \frac{\sigma_u^2}{2}\right)
\quad \text{in which} \quad
\alpha 
= \left( 1 - \frac{h-0.2}{0.8 h}\right) \sum_{a = 0}^{a_{\text{max}}} m_a f_a S^a,
\end{align}
where we used relation \eqref{eq:Arelation_E0_R0} to find the expression for $\alpha$.
Now, since $u_t - \sigma_u^2/2$ is normally distributed with mean
$-\sigma_u^2/2$ and standard deviation $\sigma_u$, we obtain,
where $\mathbf{E}$ denotes expectation,
\begin{align*}
\mathbf{E} \left\{ \text{exp} \left(u_t - \frac{\sigma_u^2}{2} \right) \right\}
= 1.
\end{align*}
Therefore, the expected number of offsprings in year $t$ becomes
\begin{align*}
\mathbf{E} \left\{ \tilde{R}_{t+1} \right\} = \frac{\tilde E_t}{\alpha}
\end{align*}
and hence the expected total number of offsprings from the single individual during its lifetime,
i.e. the basic reproduction ratio ${{\tilde R}}_{0}(F_{\text{J}},F_{\text{A}})$,
becomes
\begin{align}\label{eq:Aproof3}\tag{A17}
{{\tilde R}}_{0}(F_{\text{J}},F_{\text{A}}) = \mathbf{E} \left\{\sum_{t = 0}^{a_{\text{max}}} \tilde{{R}}_{t+1} \right\}
= \sum_{t = 0}^{a_{\text{max}}} \mathbf{E} \left\{ \tilde{{R}}_{t+1} \right\}
= \frac{1}{\alpha} \, \sum_{t = 0}^{a_{\text{max}}} \tilde E_t,
\end{align}
where $\alpha$ is given in \eqref{eq:Aproof1}.

It remains to calculate the total number of eggs produced,
i.e. the sum in the right hand side of \eqref{eq:Aproof3}.
Indeed, following (3) and (4) in the main text 
we conclude that
\begin{align*}
\sum_{t = 0}^{a_{\text{max}}} \tilde E_t
&= m_0 f_0 + m_1 f_1 S (1 - \gamma_{0}) + m_2 f_2 S^2 (1 - \gamma_{0})(1 - \gamma_{1}) + \dots \notag\\
&= m_0 f_0 + \sum_{a = 1}^{a_{\text{max}}} m_a f_a S^a \prod_{i = 0}^{a-1} (1 - \gamma_{i}).
\end{align*}
By the expression for $m_a$ in (1) in the main text 
and by $\gamma_a$ in (5) in the main text 
we find that the above expression yields
\begin{align*}
\sum_{t = 0}^{a_{\text{max}}} \tilde E_t
&= m_0 f_0 + \sum_{a = 1}^{a_{\text{max}}} m_a f_a S^a \prod_{i = 0}^{a-1} (1 - \gamma_{i})
= \sum_{a = a_{\text{mature}} + 1}^{a_{\text{max}}} f_a S^a \prod_{i = 0}^{a-1} (1 - \gamma_{i})\notag\\
&= \left(1 - F_{\text{J}}\right)^{a_{\text{mature}}}
\times \sum_{a = a_{\text{mature}} + 1}^{a_{\text{max}}} f_a S^a \left( 1 - F_{\text{A}}\right)^{a - a_{\text{mature}}}.
\end{align*}
We next 
use the relation $f_a = c s_a$ given in (1) in the main text 
to derive
\begin{align*}
\sum_{t = 0}^{a_{\text{max}}} \tilde E_t =  c \, \left( 1 - F_{\text{J}} \right)^{a_{\text{mature}}}
\times\sum_{a = a_{\text{mature}} + 1}^{a_{\text{max}}} s_a\, S^{a}  \left( 1 - F_{\text{A}}\right)^{a - a_{\text{mature}}}.
\end{align*}
Using (1) in the main text  
once more 
we see that $\alpha$ can be written
\begin{align}\label{eq:Aproof5}\tag{A18}
\alpha = c\,\left( 1 - \frac{h-0.2}{0.8 h}\right)  \sum_{a = a_{\text{mature}} + 1}^{a_{\text{max}}} s_a \,S^{a}.
\end{align}
Inserting main text eq. (3)  
and \eqref{eq:Aproof5} into \eqref{eq:Aproof3} and simplifying yield
%
\begin{align*}
{{\tilde R}}_{0}(F_{\text{J}},F_{\text{A}})
= \frac{\left(1 - F_{\text{J}}\right)^{a_{\text{mature}}} \times \sum_{a = a_{\text{mature}} + 1}^{a_{\text{max}}} s_a\, S^{a} \left( 1 - F_{\text{A}} \right)^{a - a_{\text{mature}}}}
{ \left( 1 - \frac{{{h}}-0.2}{0.8 {{h}}}\right) \times \sum_{a = a_{\text{mature}} + 1}^{a_{\text{max}}} s_a \,S^{a} }.
\end{align*}
%
This expression is identical to eq. (10) in the main text 
which gives the basic reproduction ratio for the age model.

\end{document}